\documentclass{aa}
\usepackage{graphicx}
\usepackage{txfonts}
\usepackage{natbib}
\usepackage{comment}
\usepackage{appendix}
\usepackage{longtable}
\bibpunct{(}{)}{;}{a}{}{,}
\begin{document}

\title{The HARPS search for southern extra-solar planets\thanks{Based on observations made with the HARPS instrument on the ESO 3.6 m telescope under the program IDs 072.C-0488, 082.C-0718 and 183.C-0437 at Cerro La Silla (Chile).}}

\subtitle{XXXV. Planetary systems and stellar activity of the M dwarfs GJ~3293,
 GJ~3341, and GJ~3543}

\authorrunning{Astudillo et al.}
\titlerunning{Planetary systems and stellar activity of the M dwarfs GJ~3293, 
GJ~3341, and GJ~3543}

\author{N. Astudillo-Defru \inst{1, 2}, X. Bonfils\inst{1, 2}, X. Delfosse\inst{1, 2}, D. S\'egransan\inst{3}, T. Forveille\inst{1, 2}, F. Bouchy\inst{3,4}, M. Gillon\inst{5}, C. Lovis\inst{3}, M. Mayor\inst{3}, V. Neves\inst{6}, F. Pepe\inst{3}, C. Perrier\inst{1,2}, D. Queloz\inst{3,7}, P. Rojo\inst{8}, N. C. Santos\inst{9,10}, S. Udry\inst{3}}

\institute{Univ. Grenoble Alpes, IPAG, F-38000 Grenoble, France 
  \and CNRS, IPAG, F-38000 Grenoble, France 
  \and Observatoire de Gen\`eve, Universit\'e de Gen\`eve, 
    51 ch. des Maillettes, 1290 Sauverny, Switzerland 
  \and Laboratoire d'Astrophysique de Marseille, UMR 6110 CNRS, 
    Universit\'e de Provence, 38 rue Fr\'ed\'eric Joliot-Curie, 
    13388 Marseille Cedex 13, France 
  \and Institut d'Astrophysique et de G\'eophysique, Universit\'e de Li\`ege, 
    All\'ee du 6 Ao\^ut 17, Bat. B5C, 4000 Li\`ege, Belgium 
  \and Departamento de F\'isica, Universidade Federal do Rio Grande do Norte, 
    59072-970 Natal, RN, Brazil 
  \and Cavendish Laboratory, J J Thomson Avenue, Cambridge, CB3 0HE, UK 
  \and Departamento de Astronom\'ia, Universidad de Chile, 
    Camino El Observatorio 1515, Las Condes, Santiago, Chile 
  \and Centro de Astrof\'sica, Universidade do Porto, Rua das Estrelas, 
    4150-762 Porto, Portugal 
  \and Departamento de F\'isica e Astronomia, Faculdade de Ci\^encias, 
    Universidade do Porto, Portugal 
}

\date{}

\abstract
{Planetary companions of a fixed mass induce larger amplitude reflex 
  motions around lower-mass stars, which helps make M~dwarfs excellent 
  targets for extra-solar planet searches. State of the art velocimeters 
  with $\sim$1m/s stability can detect very low-mass planets out to the 
  habitable zone of these stars. Low-mass, small, planets are abundant 
  around M~dwarfs, and most known potentially habitable planets orbit 
  one of these cool stars. 
}
{Our M-dwarf radial velocity monitoring with HARPS on the ESO 3.6m 
telescope at La Silla observatory makes a major contribution to this sample.
}
{We present here dense radial velocity (RV) time series for three M~dwarfs 
observed over $\sim5$~years: GJ~3293 (0.42M$_\odot$), GJ~3341 (0.47M$_\odot$), 
and GJ~3543 (0.45M$_\odot$). We extract those RVs through minimum $\chi^2$ 
matching of each spectrum against a high S/N ratio stack of all observed
spectra for the same star. We then vet potential orbital signals against 
several stellar activity indicators, to disentangle the Keplerian variations 
induced by planets from the spurious signals which result from rotational 
modulation of stellar surface inhomogeneities and from activity cycles.
}
{Two Neptune-mass planets - $msin(i)=1.4\pm0.1$ and $1.3\pm0.1M_{nept}$ - 
orbit GJ~3293 with periods $P=30.60\pm0.02$ d and $P=123.98\pm0.38$ d, 
possibly together with a super-Earth - $msin(i)\sim7.9\pm1.4M_\oplus$ - 
with period $P=48.14\pm0.12\;d$. A super-Earth - $msin(i)\sim6.1M_\oplus$ - 
orbits GJ~3341 with $P=14.207\pm0.007\;d$. The RV variations of GJ~3543,
on the other hand, reflect its stellar activity rather than planetary 
signals.}
{}


\maketitle
\section{Introduction}
A planet of a given mass induces a larger reflex motion on a less 
massive host star. Around the low-mass M~dwarfs, present-day observing 
facilities can consequently detect planets just a few times more massive than 
the Earth \citep{2013ApJ...766...81F, 2009A&A...507..487M}. These very low 
mass stars dominate Galactic populations by approximately 3 to 1 
\citep[e.g.][]{2010Natur.468..940V}, and most of them host planets: 
\citet{2013A&A...549A.109B} estimate that $0.88^{+0.55}_{-0.19}$ 
planets orbit each early to mid-M~dwarf with a period under 100~days,
while \citet{2013ApJ...767...95D} find that each star with effective
temperatures below $4000K$ is orbited by $0.90^{+0.04}_{-0.03}$ planets with 
radii between $0.5$ and $4R_\oplus$  and an orbital period below 50~days. 
Their high Galactic abundance and their abundant planets together 
make M~dwarfs excellent targets for planet searches. These 
stars consequently are the focus of several ongoing surveys
 - with both RV \citep[e.g. HARPS][]{2013A&A...549A.109B} and 
transit techniques \citet[e.g. MEarth][]{2008PASP..120..317N}. Several
instruments are being developed to specifically target these stars 
- e.g. SPIRou, \citet{2013sf2a.conf..497D}; CARMENES, \citet{2012SPIE.8446E..0RQ}; 
NGTS, \citet{2013EPJWC..4713002W}; Exoplanets in Transit and their Atmosphere 
(ExTrA, Bonfils et al. in prep.) - mostly in the near-infrared spectral 
range where M dwarfs are brighter and where a given photon noise can 
thus be achieved within a muchshorter integration time. 

Much interest is currently focused on discovering broadly Earth-like planets 
that orbit within the habitable zone (HZ) of their host star. The HZ zone, 
by definition, is the range of host star distances for which the incident 
stellar flux allows water on a planetary surface to remain in the liquid 
phase, and after accounting for greenhouse effects it corresponds to 
surface equilibrium temperature between 175K and 270K 
\citep{2007A&A...476.1373S}. That zone is much closer in for a 
low luminosity M~dwarf than for a brighter solar-type star: the 
orbital period for a HZ planet ranges from a week to a few months 
across the M~dwarf spectral class, compared to one year for 
the Sun-Earth system. This relaxes the ${\sim}10\;cm/s$ precision 
required to detect an Earth equivalent orbiting a Sun equivalent to
$\sim 1\;m/s$ for the same planet orbiting in the habitable zone of 
a M~dwarf. Characterizing that planet during transit, if any occurs, 
is furthermore eased considerably by the much larger planet to star 
surface ratio. The equilibrium surface temperature of a planet 
secondarily depends on the nature of its atmosphere, making planetary 
mass an important parameter as well. Bodies with $M<0.5M_\oplus$ are
expected to retain too shallow atmospheres for any water to be liquid, 
while planets with $M>10M_\oplus$ are expected to accrete a very 
thick atmosphere mainly dominated by Hydrogen and Helium 
\citep{2007A&A...476.1373S}. These considerations together make
GJ~667Cc \citep{2013A&A...553A...8D, 2013A&A...549A.109B}, 
GJ~163 \citep{2013A&A...556A.110B}, and \emph{Kepler}-186f 
\citep{2014Sci...344..277Q} some of the best current candidates
for potentially habitable planets.

Stellar activity affects habitability \citep[e.g.][]{2013A&A...557A..67V}, 
but more immediately, it can induce false-positives in planets detection. 
M~dwarfs remain active for longer than more massive stars, because 
they do not dissipate their angular momentum as fast as their more 
massive brethrens,  and stellar activity correlates strongly with 
rotation period \citep{1984ApJ...279..763N}. Additionally, lower 
mass stars are more active for a fixed rotation period 
\citep{2007AcA....57..149K}. Activity, in turn, affects measured 
stellar velocities through a number of mechanisms: stellar spots deform 
spectral lines according to their position on the stellar surface, 
the up-flowing and down-flowing regions of convective cells introduce
blue-shifted and red-shifted components to the line shapes, and stellar 
oscillations also introduce a RV jitter. Stellar activity diagnostics 
are therefore essential to filter out spurious radial velocity signals 
which can otherwise be confused with planets \citep{2007A&A...474..293B}; 

Cross-correlation with either an analogic or a numerical mask
 is widely used to extract radial velocities from spectra 
\citep{1996A&AS..119..373B}. This technique concentrates 
the information of all the lines in the mask into a very 
high signal-to-noise average line. It therefore enables a very detailed 
characterization of the line profile. Aside from the usually minor effect
of telluric absorption lines, any variation of the full-width-at-half-maximum 
(FWHM), contrast or bisector-span of the Cross-Correlation functions that 
correlate with the radial velocity variations denotes that those originate
in stellar phenomena such as spots, visible granulation density or oscillations 
\citep{2001A&A...379..279Q, 2011A&A...528A...4B, 2011A&A...527A..82D}. 
Plages or filaments on the stellar surface can additionally be detectable 
through emission in, e.g., the Ca \textrm{\small II} H\&amp;K and H$\alpha$ 
lines \citep{2011A&A...534A..30G}.

Here we present analyses of GJ~3293 and GJ~3341 for which our HARPS 
measurements  indicate the presence of planets, and for GJ~3543 where
we conclude that stellar activity more likely explains the RV variations. 
Sect.~\ref{sec:analysis_from_harps} briefly describes the observations 
and the reduction process; Sect.~\ref{sec:stellar_properties} 
discusses the properties of each star in some detail, while 
Sections~\ref{sec:GJ3293_analysis}, ~\ref{sec:GJ3341_analysis} and 
~\ref{sec:GJ3543_analysis} describe the RVs analysis and orbital 
solutions and examines stellar activity. Finally, we conclude 
in Sec.~\ref{sec:Conclusion}.

\begin{figure}[t]
\centering
\includegraphics[scale=0.45]{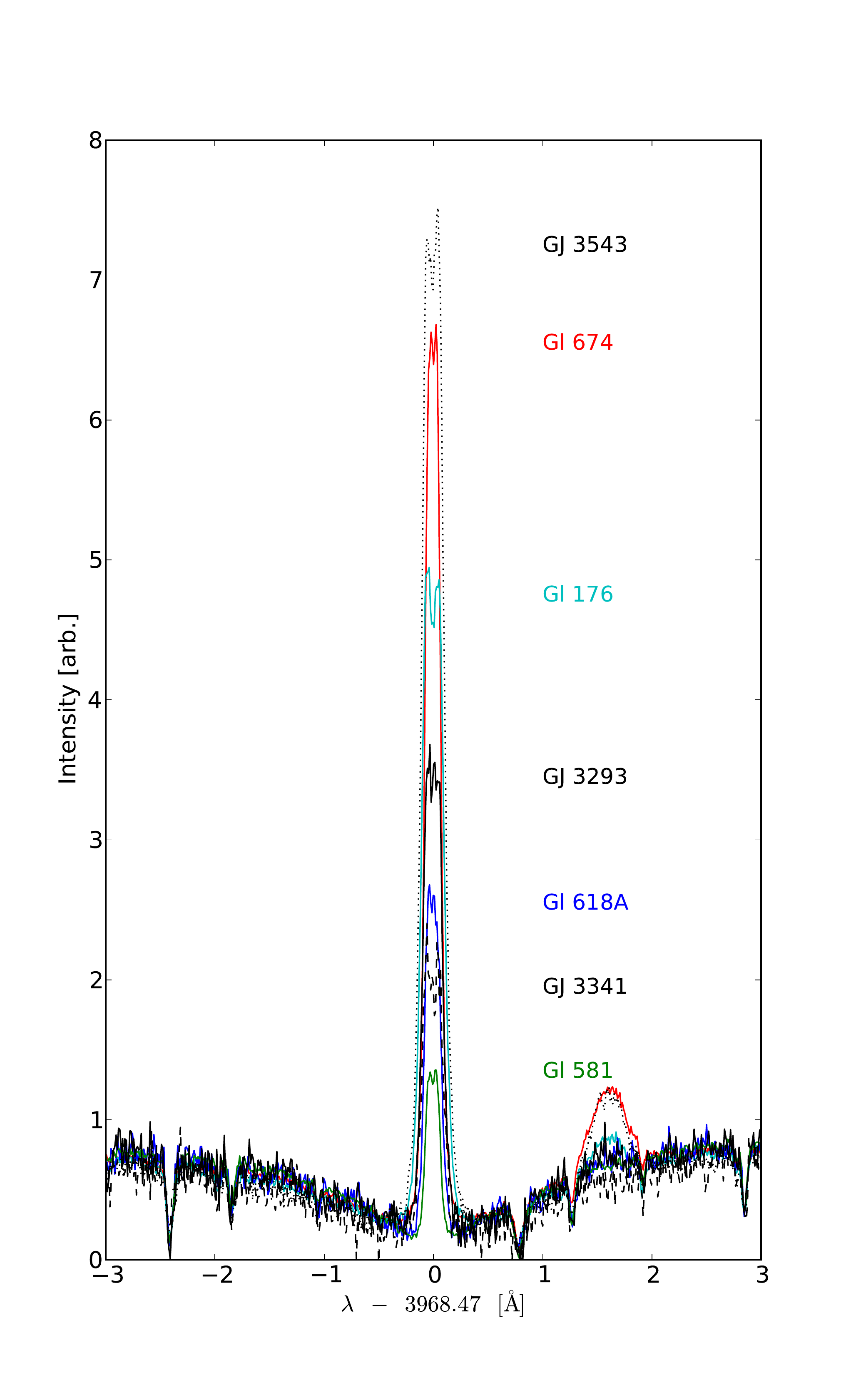}
\caption{Median spectra centered on the Ca \textrm{\small II} H line for reference stars, sorted by increasing rotation period: Gl~674 (red line, M3, $P_{rot}=35\,d$), Gl~176 (cyan line, M2.5, $P_{rot}=39\,d$), Gl~618A (blue line, M3, $P_{rot}=57\,d$), and Gl~581 (green line, M2.5, $P_{rot}=130\,d$). Median spectra for the targets of this paper, with no {\it a priori} known rotation period: GJ~3543 (black dotted line, M1.5), GJ~3293 (black full line, M2.5), and GJ~3341 (black dashed line, M2.5).}
\label{fig:activity_comparison}
\end{figure}

\section{Spectra and Doppler analysis from HARPS}
\label{sec:analysis_from_harps}
The High Accuracy Radial velocity Planets Searcher (HARPS) is a fiber-fed, 
cross-dispersed echelle spectrograph installed on the 3.6m telescope at 
La Silla observatory in Chile. The instrument diffracts the light over 
two CCDs, where 72 orders cover the 380 to 630 nm spectra range with a 
resolving power of 115,000 \citep{2003Msngr.114...20M}. HARPS stands out 
by its long term stability, ensured by a vacuum enclosure and a temperature
stabilized environment. To achieve sub-m/s precision, the spectrograph 
produces spectra for light injected through two fibers. One receives light 
from the target star and the other can be simultaneously (or not) illuminated 
with a calibration reference in order to correct instrumental drifts 
during the observations.  

The HARPS pipeline \citep{2007A&A...468.1115L} automatically reduces 
the data using nightly calibrations and measures the radial velocity 
by cross-correlation with a binary mask \citep{2002A&A...388..632P} 
which depends on the spectral type. The numerical mask for M~dwarfs 
consists in almost 10,000 holes, placed on spectral lines selected
for their large amount of Doppler information. The whole procedure 
completes shortly after the end of each exposure.

The visual band spectra of the coolest stars contain a very large numbers of 
overlapping molecular features with essentially no continua. Under
such circumstances, a binary mask makes sub-optimal use of the
available Doppler information. In this study, we therefore recomputed 
RVs from the order by order spectra extracted by the HARPS pipeline. 
For each target, we used the RVs measured by the HARPS pipeline for 
the individual spectra together with the corresponding barycentric 
correction to align all spectra to the frame of the Solar System barycenter. 
This aligns the stellar lines, while the telluric features are 
shifted by minus the barycentric velocity of each epoch. We then compute 
the median of these spectra to produce a high SNR template
spectrum for each target. At that stage, we produce a template of the 
telluric absorption spectrum, by computing the median of the residuals 
(aligned in the laboratory reference frame) of subtracting the high SNR 
template from the individual spectra. We then use this telluric spectrum 
to produce an improved stellar template, by repeating its construction 
with the now known telluric lines masked out. This process can in principle
be iterated, but we found that it effectively converge after the first 
iteration. Finally, we measure a new radial velocities by minimizing 
the chi-squared of the residuals between the observed spectra and 
shifted versions of the stellar template, with all spectral elements 
contaminated by telluric lines masked out \citep[e.g.][Astudillo et al. 
in prep.]{1997MNRAS.284..265H, 2006MNRAS.371.1513Z}. Astudillo et al. 
(in prep.) will provide a detailed description of the algorithm implementation 
and will characterize its performance.

Our observation strategy is described in detail in 
\citet{2013A&A...549A.109B}, and only summarized for convenience here. 
We chose to observe without illuminating the reference fiber, as we 
only targeted a $\sim1\;ms^{-1}$ precision; this choice provides
clean observations of the Ca \textrm{\small II} H\&K lines for later 
stellar activity analysis, which is particularly important for M dwarfs. 
We hence made use of wavelength calibrations acquired before the beginning of 
the night. The exposure time was 900s for all frames. This is adequate for 
$0.80\;ms^{-1}$ precision for visual magnitudes between 7 and 10, but the 
velocities of the fainter stars which we discuss here have significantly 
higher photon noise errors.

\section{Stellar properties of GJ 3293, GJ 3341, and GJ 3543}
\label{sec:stellar_properties}

GJ~3293 (LHS~1672), GJ~3341 (LHS~1748), and GJ~3543 (L~749-34) are high 
proper motions early M~dwarfs (M2.5, M2.5 and M1.5, respectively). We 
used the  $BC_K$ bolometric correction of \citet{2001ApJ...548..908L} 
and the photometric distance of \citet{1991adc..rept.....G} to compute 
their luminosity. We also estimated the effective temperature 
($T_{eff}$), stellar radius, 
and luminosity from the $V-K$ color and metallicity relationship of 
\citet{2012ApJ...757..112B}; the two luminosities agree well for the 
three targets. We derived the stellar metallicities - and $T_{eff}$, 
for comparison - from our spectra using the methods of 
\citet{2014arXiv1406.6127N}; the two determinations of $T_{eff}$ agree to
better than their error bars for all three stars, and we only quote the 
\citet{2012ApJ...757..112B} value. The masses were computed using 
the \citet{2000A&A...364..217D} K-band mass versus absolute magnitude 
relation. We calculated the UVW space motions with the 
\citep{1987AJ.....93..864J} orientation convention, and assign kinematic 
populations following \citet{1992ApJS...82..351L}. We used the proper 
motion and distance to compute the secular radial acceleration 
$dv/dt$ \citep{2003A&A...403.1077K}, from which we corrected the 
radial velocities. Following \citet{2007A&A...476.1373S}, we adopt  
recent Venus and early Mars criteria for the inner ($HZ_{In}$) and outer 
($HZ_{Out}$) edges of the Habitable Zone. Table~\ref{tab:targetsProperties} 
summarizes the properties of the three targets.

GJ~3293 is located in the Eridanus constellation and $18.2\pm2.6$~pc 
\citep{1991adc..rept.....G} away from the Sun. Its Galactic velocity 
parameters, $U=-27.3\pm17.1\, kms^{-1}$, $V=-25.9\pm6.6\, kms^{-1}$, and 
$W=-22.2\pm23.1\,kms^{-1}$, leave its kinematic population uncertain 
in part due to the large uncertainty on its photometric distance; GJ~3293 
could belong either in the young disk or the young-old disk population. 
Its close to Solar metallicity ([Fe/H]=0.02) suggests that it is
part of the young disk, but is consistent with either option.

GJ~3341 is located in the Columba constellation at a distance of 
$23.2\pm0.7$ pc \citep{2010AJ....140..897R}. Its proper motion, distance, 
and systemic velocity ($\gamma=47.803\pm0.003$) result in 
$U=52.5\pm0.6\, kms^{-1}$, $V=-52.0\pm0.8\, kms^{-1}$, and 
$W=24.4\pm3.2\,kms^{-1}$. This formally makes GJ~3341 fits a 
young-old disk member. 
 
GJ~3543 is located in the Hydra constellation and at 
$12.5\pm2.0$~pc from the Sun \citep{1991adc..rept.....G}. 
Its space motions components $U=23.8\pm11.3\, kms^{-1}$, 
$V=-9.0\pm2.0\, kms^{-1}$, and $W=-2.7\pm1.7\,kms^{-1}$ place
GJ~3543 in the young disk box while its metallicity ([Fe/H]=-0.13) 
is somewhat low for the Galactic young disk.

\begin{table}[t]
\centering
\begin{tabular*}{\hsize}{@{\extracolsep{\fill}}llll}

\hline
\noalign{\smallskip}
& GJ~3293 & GJ~3341 & GJ~3543\\
\noalign{\smallskip}
\hline\hline
\noalign{\smallskip}
Spectral Type & M2.5 & M2.5  & M1.5 \\
$\alpha$ (J2000) & $04^h 28^m 35.6^s$ & $05^h 15^m 46.7^s$ & $09^h 16^m 20.7^s$ \\
$\delta$ (J2000) & $-25^\circ 10^\prime 16^{\prime\prime}$ & $-31^\circ 17^\prime 46^{\prime\prime}$ & $-18^\circ 37^\prime 33^{\prime\prime}$ \\
V$^{(1)}$ & 11.962 & 12.080 & 10.739\\
J$^{(2)}$ & $8.362\pm0.024$ & $8.592\pm0.020$ & $7.351\pm0.021$\\
H$^{(2)}$ & $7.749\pm0.038$ & $7.990\pm0.049$ & $6.759\pm0.047$\\
K$^{(2)}$ & $7.486\pm0.033$ & $7.733\pm0.023$ & $6.492\pm0.024$\\
$\pi\; [mas]^{(3, 4, 3)}$& $55\pm9$ & $43.18\pm1.40$ & $80\pm15$\\
$M_V$ & $10.66\pm0.31$ & $10.26\pm0.07$ & $10.25\pm0.34$ \\
$M_K$ & $6.19\pm0.31$ & $5.91\pm0.07$ & $6.01\pm0.34$ \\
$BC_K$ & $2.71\pm0.08$ & $2.68\pm0.06$ & $2.68\pm0.06$ \\
L [L$_{Sun}$]$^{(6)}$ & 0.022 & 0.029 & 0.026\\
M [M$_\odot$]$^{(5)}$ & 0.42 & 0.47 & 0.45\\
R [R$_\odot$]$^{(7)}$ & $0.404\pm0.027$ & $0.439\pm0.027$ & $0.432\pm0.027$\\
T$_{eff}$ [K]$^{(7)}$ & $3466\pm49$ & $3526\pm49$ & $3524\pm49$\\
Fe/H$^{(8)}$ & $0.02\pm0.09$  & $-0.09\pm0.09$ & $-0.13\pm0.09$\\
$\mu_\alpha \;[mas/yr]^{(9)}$ & $-87\pm5$  & $504\pm5$ &  $-314.6\pm3.4$\\
$\mu_\delta \;[mas/yr]^{(9)}$ & $-475\pm5$ & $243\pm5$ & $148.0\pm3.3$\\
$dv_r/dt \; [m/s/yr]^{(10)}$ & $0.097\pm0.018$ & $0.167\pm0.008$ & $0.035\pm0.007$\\
$HZ_{In}\;[AU]^{(11)}$ & 0.112 & 0.129 & 0.124\\
$HZ_{Out}\;[AU]^{(11)}$ & 0.305 & 0.350 & 0.335 \\

\noalign{\smallskip}
\hline
\end{tabular*}
\caption{(1) \citet{2012yCat.1322....0Z}; (2) \citet{2003yCat.2246....0C}; (3) \citet{1991adc..rept.....G}; (4) \citet{2010AJ....140..897R}; from (5) \citet{2000A&A...364..217D}, (6) \citet{2001ApJ...548..908L}, (7) \citet{2012ApJ...757..112B} and (8) \citet{2014arXiv1406.6127N} relation ships; (9) \citet{2003ApJ...582.1011S}; (10) \citet{2003A&A...403.1077K}; (11) \citet{2007A&A...476.1373S}}
\label{tab:targetsProperties}
\end{table}

\section{Radial velocities of GJ 3293}
\label{sec:GJ3293_analysis}

The 145 RV measurements of GJ~3293 span 1514~d. Their $\sigma_e=7.69\,ms^{-1}$ 
dispersion is much larger than the average Doppler uncertainty 
$\langle\sigma_i\rangle=1.76\,ms^{-1}$, which 
represents the weighted arithmetic mean of the estimated photon noise 
\citep{2001A&A...374..733B} and instrumental errors. 
Both an F-test with $F=\sigma_e^2 /\langle \sigma_i\rangle^2$ and a 
$\chi^2$ test for a constant model given $\langle\sigma_i\rangle$
return negligible probabilities ($<10^{-9}$) that the photon noise 
combined with wavelength calibration and guiding uncertainties 
explain the measured dispersion.

We thus looked for periodicity with floating-mean periodograms, with 
a periodogram normalization choice where 1 stands for a perfect fit 
of a sine wave to the data and 0 points to no improvement over a 
constant model \citep{2009A&A...496..577Z}. Besides the commonly used 
$1\%$ False Alarm Probability (FAP) confidence level, we plot values 
covering $68.3\%$, $95.4\%$, and $99.7\%$ of the periodogram power 
distributions, equivalent to 1$\sigma$ ($31.7\%$ FAP), 2$\sigma$ 
($4.6\%$ FAP), and 3$\sigma$ confidences ($0.3\%$ FAP).

\begin{figure}[t]
\centering
\includegraphics[scale=0.45]{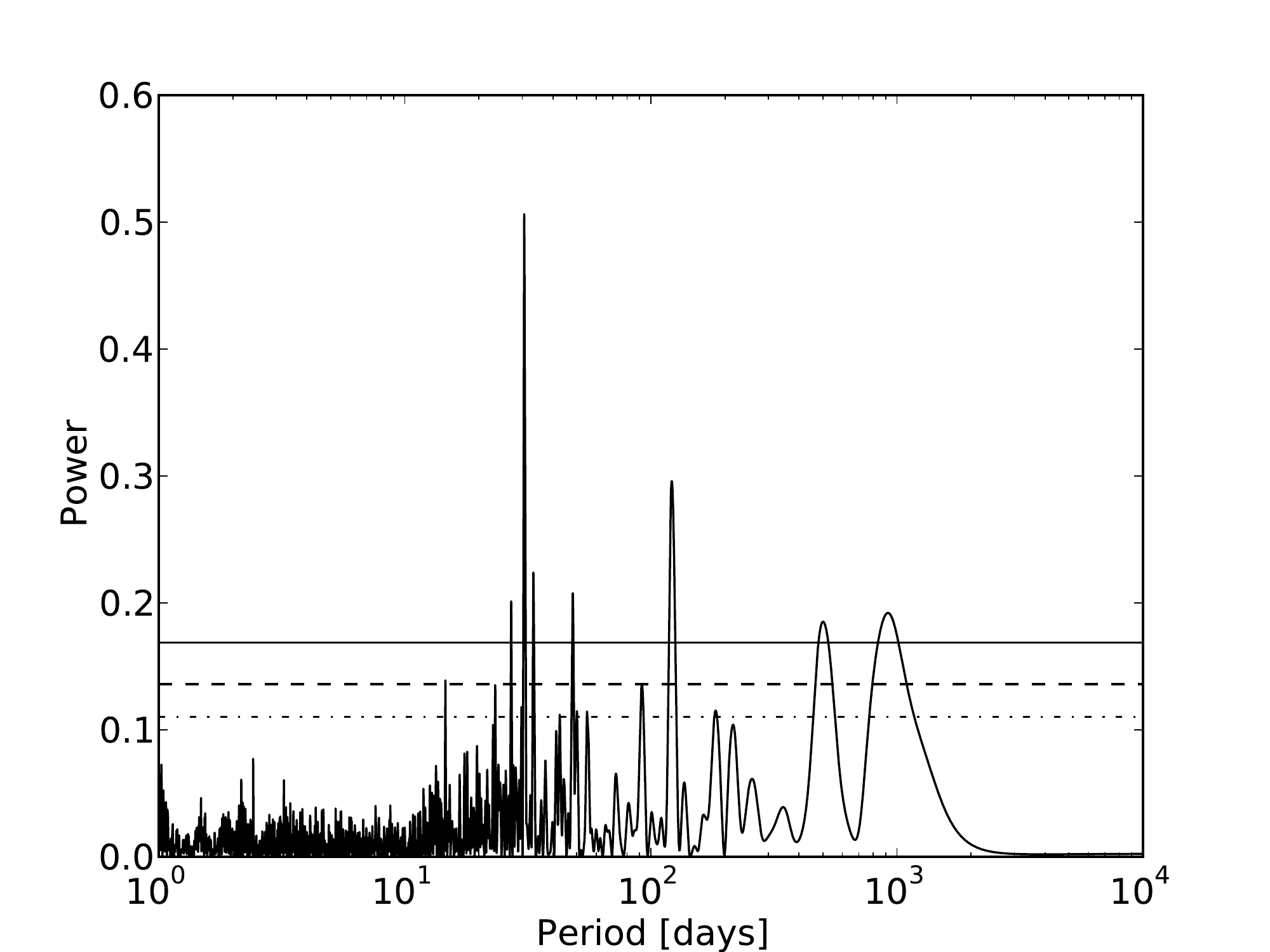}
\includegraphics[scale=0.45]{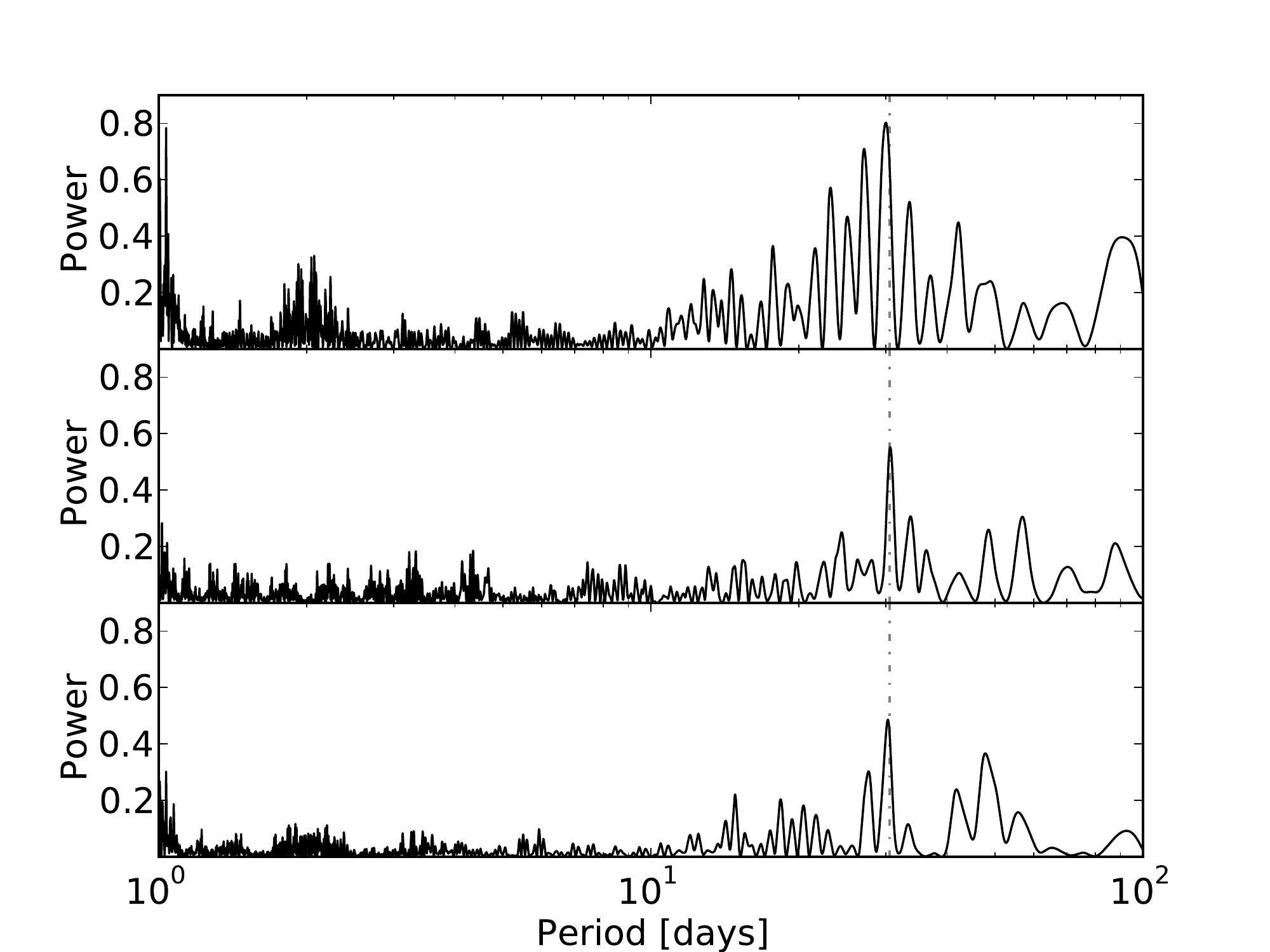}
\caption{\emph{Top panel:} Periodogram of the GJ~3293 RVs. The solid line, 
dashed line, and dashed dotted line represent the $0.3\%$, $4.6\%$, and 
$31.7\%$ FAP levels, corresponding to $3\sigma$, $2\sigma$, $1\sigma$ 
confidence, respectively. \emph{Bottom panel:} Periodograms for 
epochs 2008-2009 (34 measurements, first row), 2010-2011 (52 measurements, 
second row), and 2012-2013 (59 measurements, third row); the stability of 
the 30 d signal (dashed dotted vertical) is clear.}
\label{fig:GJ3293_RV_PeriodogramOBS}
\end{figure}

Fig.~\ref{fig:GJ3293_RV_PeriodogramOBS} shows the periodogram of the GJ~3293 
time series and reveals a clear power excess around $P=30.6$ d, with 
$p_{max}=0.51$. Additional peaks above the $0.3\%$ FAP ($p=0.17$) appear 
at 121.6, 33.3, 48.2, 27.1, 919.5, and 500.9~d, with powers of 0.30, 
0.22, 0.21, 0.20, 0.19, and 0.19, respectively. To further evaluate
the confidence on the  $P=30.6\;d$ signal given our measurement errors
and sampling, we generated $1,000$ synthetic datasets by rearranging the 
radial velocities and holding the dates fixed. None of the periodograms 
generated for these bootstraped had maximum power above 0.3. The FAP
on the 30.6 d signal, with 0.51 power, is therefore well below 
1/1,000. The \citet{1986ApJ...302..757H} prescription for periodogram 
interpretation gives $FAP(30.6d)=2.8 \times 10^{-19}$, and the 30.6~d peak 
is well above any of the considered confidence levels.

We used \emph{yorbit} (S\'egransan et al. in prep) to adjust keplerian 
orbits with an MCMC algorithm. Without any prior on the orbit, this 
converged to a solution with period $P=30.565\pm0.024\,d$, 
eccentricity $e=0.158\pm0.082$, and semi-amplitude $K_1=8.87\pm0.83\,ms^{-1}$. 
This solution reduces the rms dispersion of the residuals to 
$\sigma_e=5.34\,ms^{-1}$ and the reduced chi-square to 
$\chi_\nu^2=9.28\pm0.37$. Given a M=0.42M$_\odot$ stellar mass 
(with $10\%$ uncertainty), the minimum mass for the planet is 
$m\,sin(i)=1.4\pm0.1M_{nept}$. Table~\ref{tab:GJ3293_k1} 
summarizes the orbital and derived parameters. The ratio of the 
eccentricity (e) to its uncertainty ($\sigma_e$) is $e/\sigma_e<2.49$, 
and therefore below the usual thresholds for significant 
eccentricity\footnote{$\varepsilon_{95}/\mu=3.34$ for the eccentricity 
upper limit, where $\mu=\sigma_e$ and $\alpha(\%)=5$ for the detection 
threshold - using \citet{2013A&A...551A..47L} nomenclature.} 
\citep{2013A&A...551A..47L}. We adopt the eccentricity that 
\emph{yorbit} converged to when analysing the residuals for additional 
signals, but its small value makes that choice unimportant.

\begin{table}[t]
\caption{One-keplerian fit for GJ~3293}
\label{tab:GJ3293_k1}
\centering
\begin{tabular}{l l}
\hline
\noalign{\smallskip}
& GJ~3293b \\
\noalign{\smallskip}
\hline\hline
\noalign{\smallskip}
P [d] & $30.57\pm0.02$\\
$T_0$ [JD-2400000] & $55661.6\pm2.3$\\
$\omega$ [deg] & $169.4\pm26.7$\\
e & $0.16\pm0.08$\\
$K_1$ [$ms^{-1}$] & $8.9\pm0.8$\\
$m\, sin(i)$ [$M_{nept}$] & $1.4\pm0.1$\\
a [AU] & 0.1433\\
\noalign{\smallskip}
\hline
\noalign{\smallskip}
$N_{meas}$&145\\
Span [d]& 1514\\
$\langle\sigma_i\rangle\,[ms^{-1}]$ & 1.76\\
$\sigma_e \, [ms^{-1}]$ & 5.34\\
$\chi_\nu^2$&9.28\\
\noalign{\smallskip}
\hline
\end{tabular}
\end{table}

\begin{figure}[t]
\centering
\includegraphics[scale=0.45]{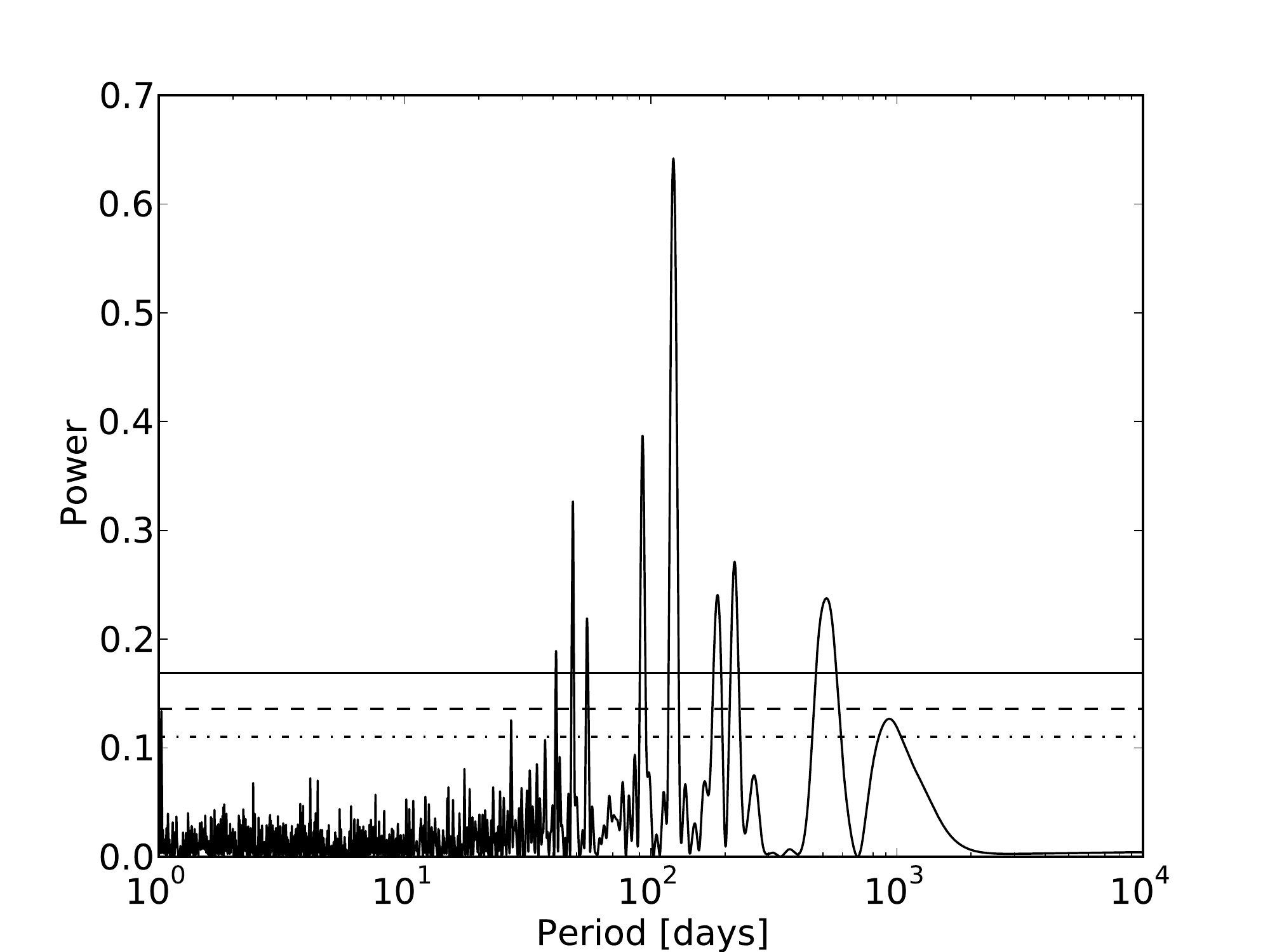}
\includegraphics[scale=0.45]{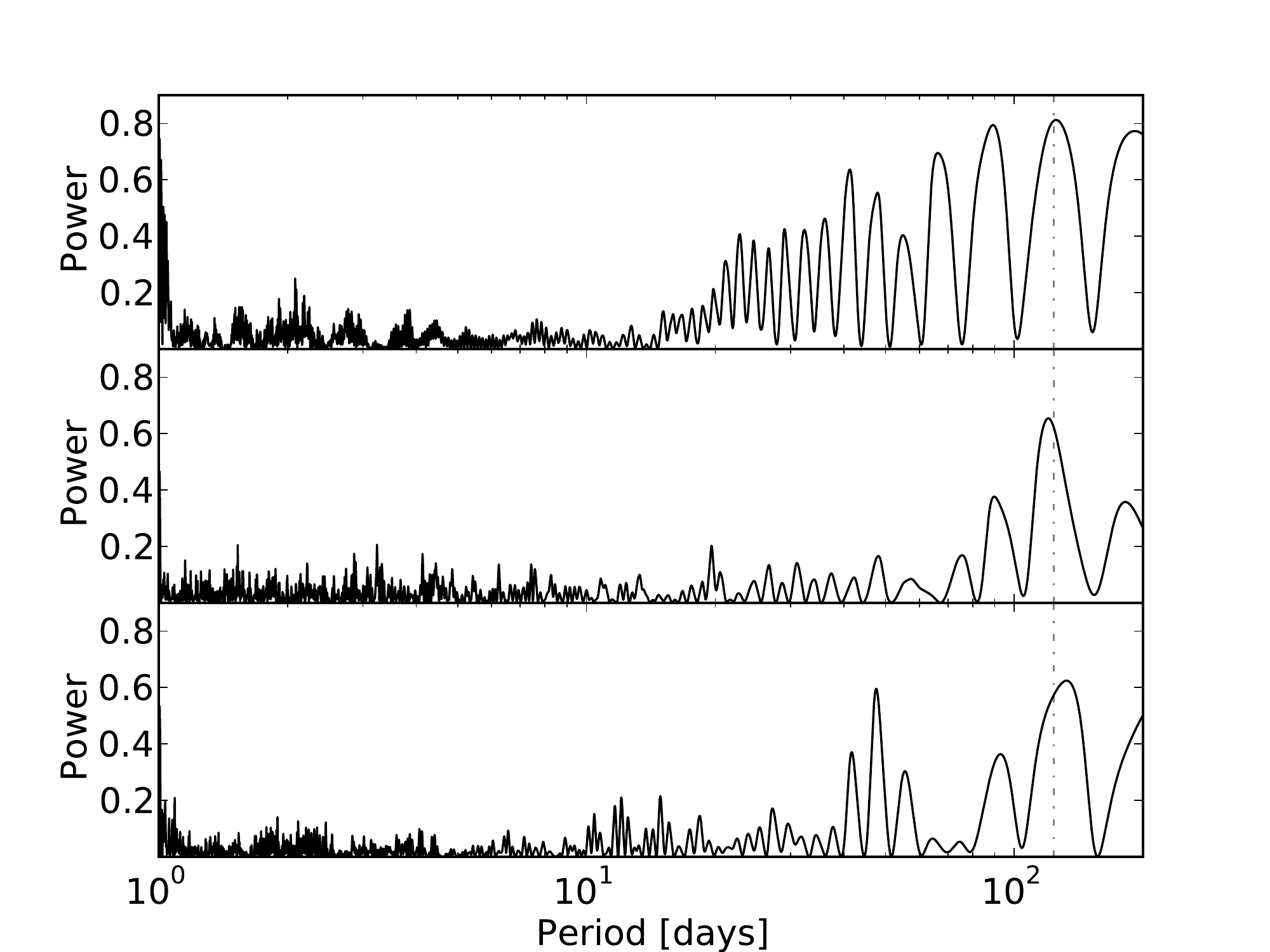}
\caption{\emph{Top panel:} Periodogram of the residuals from the 
subtraction of the first keplerian (P=30.6 d). The horizontal lines 
have the same meaning as in Fig.~\ref{fig:GJ3293_RV_PeriodogramOBS}. 
\emph{Bottom panel:} Periodogram We split for subsets of the observational 
epochs, defined in the caption to Fig.~\ref{fig:GJ3293_RV_PeriodogramOBS}; 
in spite of the poor sampling of some of the subsets for a 123~d period, 
all show a peak around this period.}
\label{fig:GJ3293_RV_PeriodogramOMCk1}
\end{figure}

Many of the peaks in the top panel of 
Fig~\ref{fig:GJ3293_RV_PeriodogramOBS} have no 
counterpart in the periodogram of the residuals of the subtraction of 
the 30.6~d signal (Fig.~\ref{fig:GJ3293_RV_PeriodogramOMCk1}) 
and therefore represent no more than aliases of that signal. A peak 
around 123.4 d dominates this periodogram of the residuals, with 
$p_{max}=0.64$. A bootstrap test with one thousand (1,000) iterations produced
no signal above 0.3, and the FAP of the dominant peak is therefore
well under $10^{-3}$. The prescription of \citet{1986ApJ...302..757H}
evaluates the FAP to $3.3 \times 10^{-29}$ for the 123.4 d peak. 
Other peaks above a $0.3\%$ FAP ($p=0.17$) occur at periods 92.1, 
48.2, 218.8, 186.5, 517.2, 55.0, and 41.2 d, and with powers of 
0.39, 0.33, 0.27, 0.24, 0.24, 0.22, and 0.19. None of those is 
sufficiently strong that confusing the 123.4 d signal for one
of its aliases would be an issue. We used \emph{yorbit} to model 
the RVs with two Keplerian signals, again with no prior on the
orbital parameters. The parameters of the first Keplerian are
essentially unchanged from the one-Keplerian fit, and the second 
has a period $P=123.76\pm0.30\,d$, eccentricity $e=0.331\pm0.057$, and 
semi-amplitude $K_1=6.430\pm0.423\,ms^{-1}$, which correspond to
a minimum planetary mass of $m\,sin(i)=1.5\pm0.1M_{nept}$. 
Table~\ref{tab:GJ3293_k2} summarizes the parameters of the two keplerians. 
The dispersion and the reduced chi-square decrease to 
respectively $\sigma_e=2.86\,ms^{-1}$ and $\chi_\nu^2=2.76\pm0.20$. 
An F-test of this new  $\sigma_e^2$ against the average internal
errors $\langle\sigma_i\rangle^2$ and a $\chi^2$ against a constant 
model respectively return $P(F)=5.7 \times 10^{-9}$ and $P(\chi^2)<10^{-9}$, 
so there remains significant dispersion above the internal errors.

\begin{table}[t]
\caption{Fit for two keplerian orbits for GJ~3293}
\label{tab:GJ3293_k2}
\centering
\begin{tabular}{l l l}
\hline
\noalign{\smallskip}
& GJ~3293b & GJ~3293c\\
\noalign{\smallskip}
\hline\hline
\noalign{\smallskip}
P [d] & $30.60\pm0.01$ & $123.75\pm0.30$\\
$T_0$ [JD-2400000] & $55638.7\pm2.6$ & $55687.8\pm3.8$\\
$\omega$ [deg] & $-103.0\pm31.3$ & $-21.4\pm14.2$\\
e & $0.07\pm0.04$ & $0.33\pm0.06$\\
$K_1$ [$ms^{-1}$] & $9.6\pm0.4$ & $6.4\pm0.4$\\
$m\, sin(i)$ [$M_{nept}$] & $1.5\pm0.1$ & $1.5\pm0.1$\\
a [AU] & 0.1434 & 0.3640\\
\noalign{\smallskip}
\hline
\noalign{\smallskip}
$N_{meas}$ & 145\\
Span [d] & 1514\\
$\langle\sigma_i\rangle\,[ms^{-1}]$ & 1.76\\
$\sigma_e \, [ms^{-1}]$ & 2.86\\
$\chi_\nu^2$ & 2.76 \\
\noalign{\smallskip}
\hline
\end{tabular}
\end{table}

\begin{figure}[t]
\centering
\includegraphics[scale=0.45]{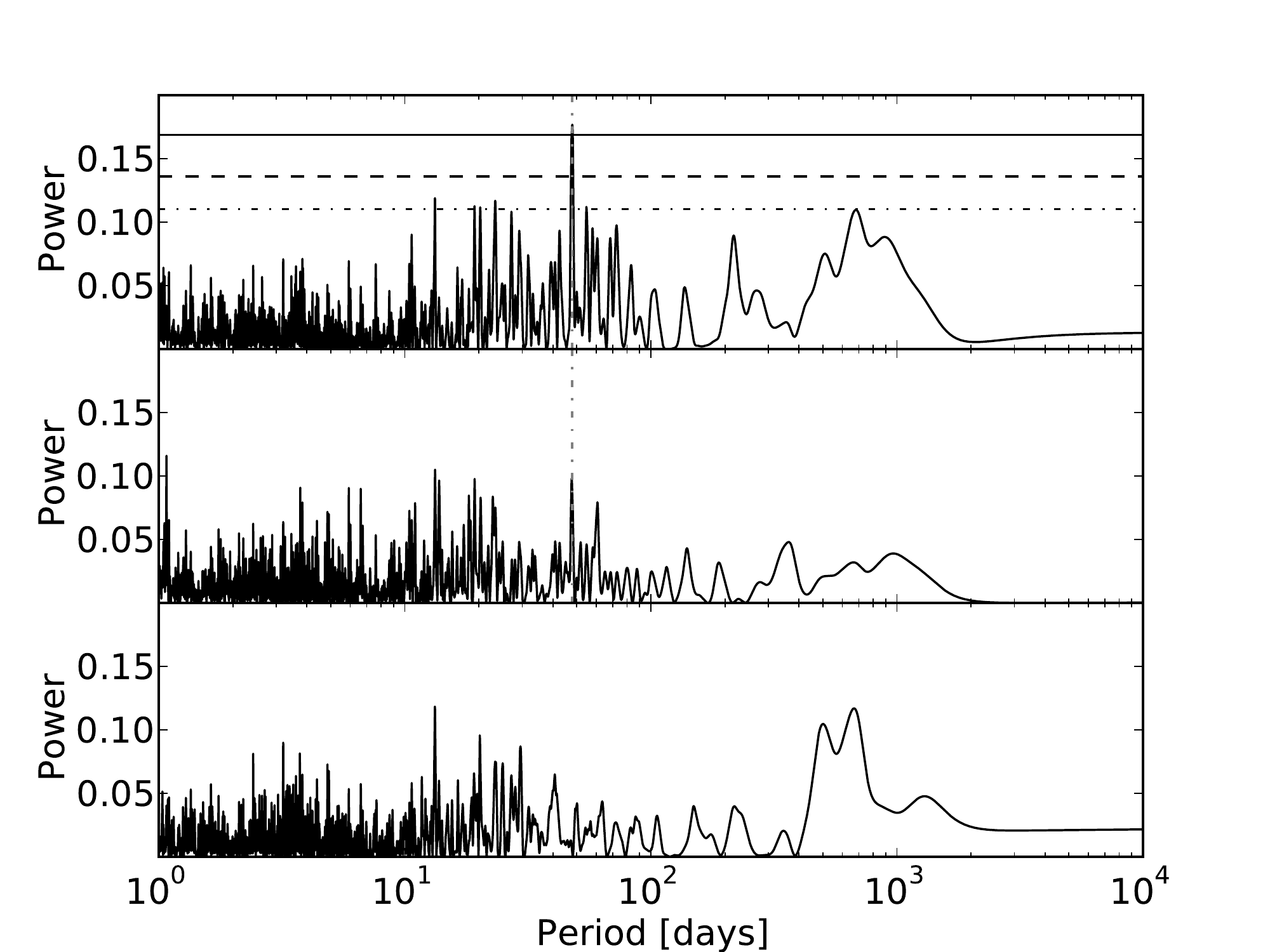}
\includegraphics[scale=0.45]{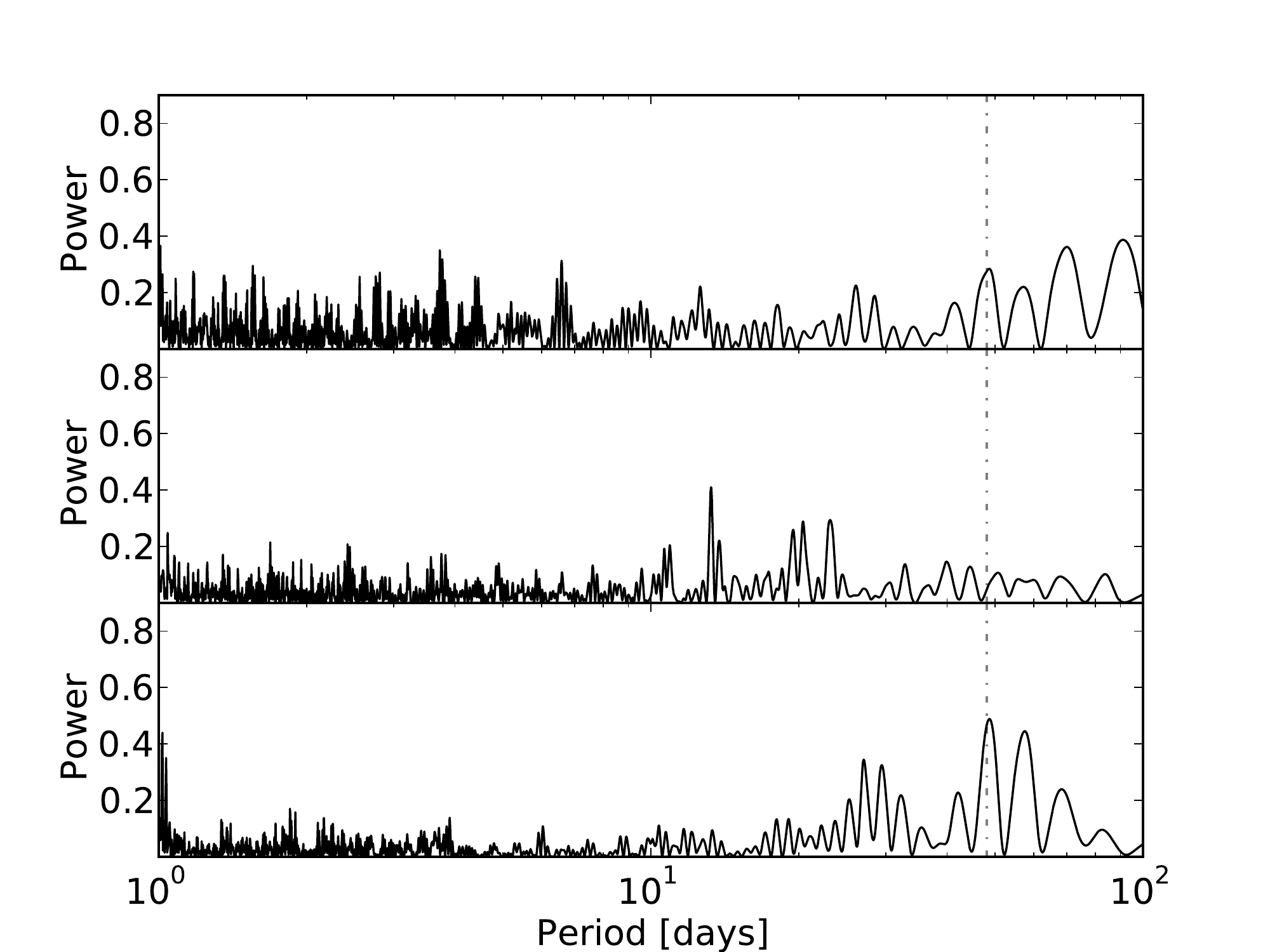}
\caption{\emph{Top panel}. \emph{First row:} Periodogram of the residuals 
of the subtraction of two Keplerians with periods of 30.6 and 123.6~d,
dominated by a 48~d peak. The horizontal lines represent FAP levels,
as described in the captions to Fig.~\ref{fig:GJ3293_RV_PeriodogramOBS}. 
\emph{Second row}: Periodogram of the residuals after subtraction of three 
Keplerians with periods of 30.6, 123.6, and 439 d. The vertical 
dashed-dotted line marks a $P=48$~d period. \emph{Third row}:  Periodogram 
of the residuals after subtraction of Keplerians with periods of 30.6, 
123.6, and 48 d. \emph{Bottom panel}. Periodogram of the residuals 
after subtraction of the 30.6 and 123.6 d signals for each subset 
of epochs described in the captions to Fig.~\ref{fig:GJ3293_RV_PeriodogramOBS}. 
The 48 d signal is not seen for the 2010-2011 observational epochs, 
which have poor sensitivity to that period range because their sampling 
happens to concentrate around just two phases.}
\label{fig:GJ3293_RV_PeriodogramOMCk2}
\end{figure}

\begin{figure}[t]
\centering
\includegraphics[scale=0.45]{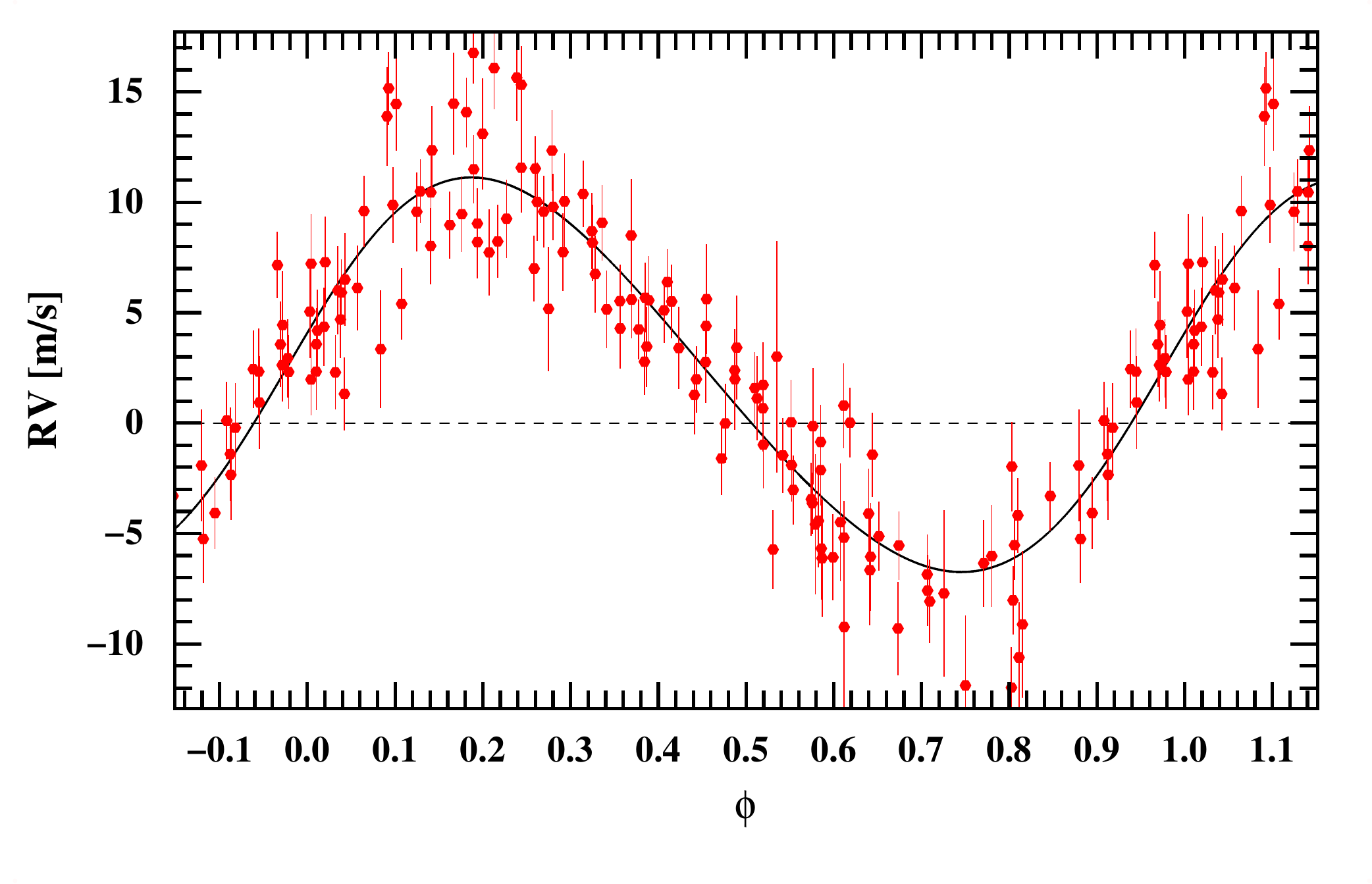}
\includegraphics[scale=0.45]{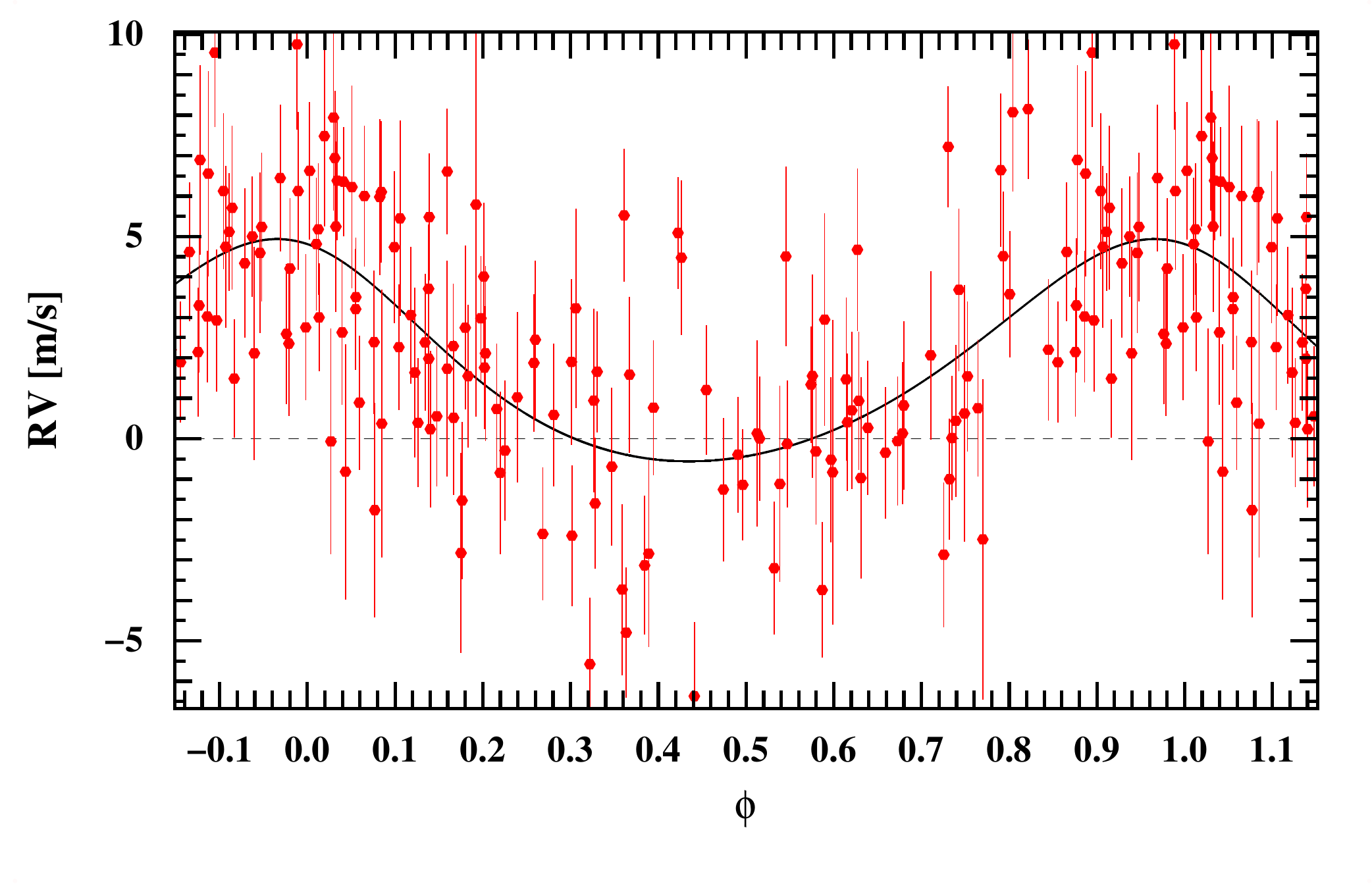}
\includegraphics[scale=0.45]{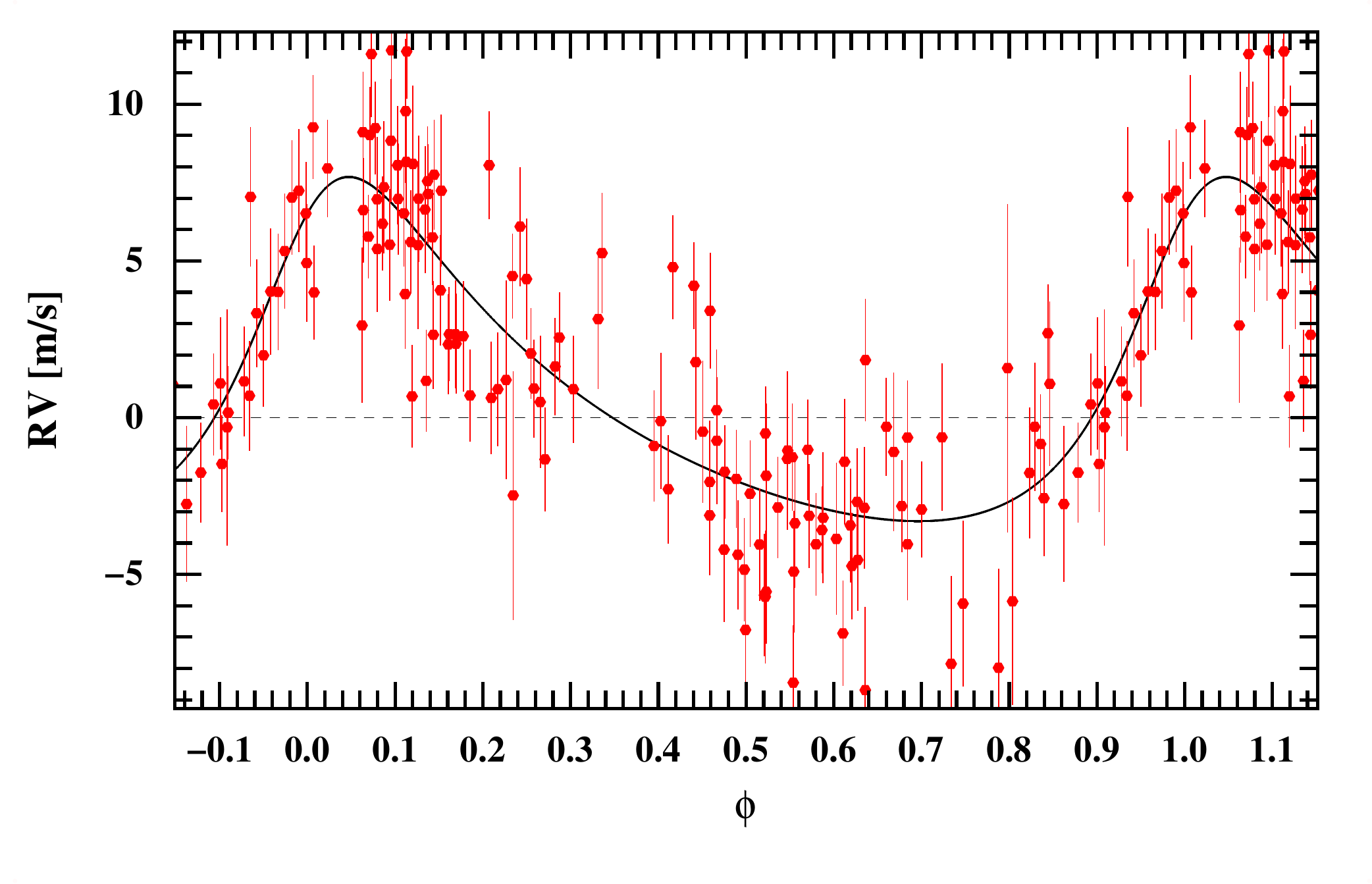}
\caption{Radial velocities phased for each signal. P=30.59~d top panel; P=48.07~d middle panel; and P=123.79~d bottom panel.}
\label{fig:GJ3293_RV_phase}
\end{figure}

A single peak dominates the periodogram of the residuals of the 
subtraction of the two keplerians (Fig.~\ref{fig:GJ3293_RV_PeriodogramOMCk2}), 
implying that the other strong peaks in Figs~\ref{fig:GJ3293_RV_PeriodogramOBS} 
and~\ref{fig:GJ3293_RV_PeriodogramOMCk1} are aliases of the 30.6 and 123.8~d
signals. This peak at 48~d has power $p_{max}=0.18$, which corresponds to 
a $0.15\%$ FAP ($p=0.15$ corresponds to a $1\%$ FAP and $p=0.17$ to a 
$3\sigma$ confidence level). An unconstrained search for a three-Keplerian
solution with \emph{yorbit} converged to the two Keplerians described above 
plus a highly eccentric ($e=0.925\pm0.022$) Keplerian with a period of 
439~d period. The third orbit crosses the other two, making the 
solution almost certainly unstable on very short time scales, and therefore 
unphysical. Spurious highly eccentric orbits are favored when noise 
becomes significant and/or sampling is poor, with the highest velocity
excursions typically found at the worst sampled phases of the orbit. 
The periodogram of the residuals of that unphysical solution still has 
a 48~d peak, but with much reduced power ($p=0.10$, 0.63 FAP, 
middle panel of Fig.~\ref{fig:GJ3293_RV_PeriodogramOMCk2}). This indicates 
that our sampling couples signals at periods of 48 and 439~d, but 
incompletely. We therefore constrained the period of the third Keplerian
to the [2, 100]~d range, to avoid convergence on the spurious longer
period eccentric solution. This converged to a Keplerian with 
$P=48.072\pm0.120\,d$, $e=0.190\pm0.134$, and $K_1=2.515\pm0.393\,ms^{-1}$, 
which corresponds to a minimum mass of $m\,sin(i)=7.9\pm1.4M_\oplus$, plus 
the two keplerians with periods of 30.6 and 123.4 d. Following 
\citet{2013A&A...551A..47L}, $e_b/\mu<2.49$ and the eccentricity 
therefore remains below the detection threshold. 
Figure~\ref{fig:GJ3293_RV_phase} shows the keplerian solution. The dispersion 
is $\sigma_e=2.45\,ms^{-1}$ and the reduced chi-square is  
$\chi_\nu^2=2.11\pm0.18$. An F-test of this $\sigma_e^2$ against 
$\langle\sigma_i\rangle^2$ yields a $P(F)=4.3 \times 10^{-5}$ probability
that this would occur by chance. The RVs therefore vary by significantly more
than expected from their known measurement errors. Possible explanations
include additional companions, stellar activity, or a non-Gaussian or
underestimated noise. The periodogram of the residuals of the 3-Keplerians 
solution (bottom panel of Fig.~\ref{fig:GJ3293_RV_PeriodogramOMCk2} has
no peak above a 12\% FAP ($11.8\%$ at 13.3 d and $11.7\%$ at 669.6 d). 
Our final solution (Table~\ref{tab:GJ3293_k3}) additionally includes a 
quadratic drift, which improves the residuals by a formally significant 
amount and suggests a possible component at a wider separation. 

\begin{table}[h]
\caption{Fit for three keplerian orbits plus a quadratic drift for GJ~3293}
\label{tab:GJ3293_k3}
\centering
\begin{tabular}{l l l l}
\hline
\noalign{\smallskip}
& GJ~3293b & GJ~3293(c) & GJ~3293d\\
\noalign{\smallskip}
\hline\hline
\noalign{\smallskip}
P [d]                 & $30.60\pm0.02$  & $48.14\pm0.12$  & $123.98\pm0.38$\\
$T_0$ [JD-2400000]       & $55640.3\pm1.9$ & $55643.3\pm6.0$ & $55684.9\pm3.7$\\
$\omega$ [deg]           & $-77.7\pm24.2$  & $17.3\pm46.3$   & $-38.0\pm14.1$ \\
e                        & $0.09\pm0.04$   & $0.16\pm0.13$   & $0.37\pm0.06$  \\
$K_1$ [$ms^{-1}$]         & $8.9\pm0.4$     & $2.7\pm0.4$     & $5.5\pm0.4$    \\
$m\, sin(i)$ [$M_{nept}$] & $1.4\pm0.1$     & $0.5\pm0.1$     & $1.3\pm0.1$ \\
a [AU]                   & 0.1434          & 0.1939          & 0.3644          \\
\noalign{\smallskip}
\hline
\noalign{\smallskip}
$\gamma \; [kms^{-1}]$ & $13.297\pm0.018$ \\
$\dot \gamma \; [ms^{-1}yr^{-1}]$ & $-0.103\pm0.002$ \\
$\ddot\gamma \; [ms^{-1}yr^{-2}]$ & $0.325 +- 0.001$ \\
$N_{meas}$ & 145\\
Span [d] & 1514\\
$\langle\sigma_i\rangle\,[ms^{-1}]$ & 1.76\\
$\sigma_e \, [ms^{-1}]$ & 2.41\\
$\chi_\nu^2$ & 2.07\\
\noalign{\smallskip}
\hline
\end{tabular}
\end{table}

\subsection{Stellar activity}
\label{sec:GJ3293_activity}
We computed periodograms for the FWHM, bissector span, and contrast of the 
CCF, as well as for the S and H$\alpha$ indices, to investigate whether
some of the periodicities are attributed to stellar activity. 
We also look for correlations between these activity indicators and 
the radial velocities and their residuals after subtracting subsets
of the Keplerian orbits.

\begin{figure}[t]
\centering
\includegraphics[scale=0.45]{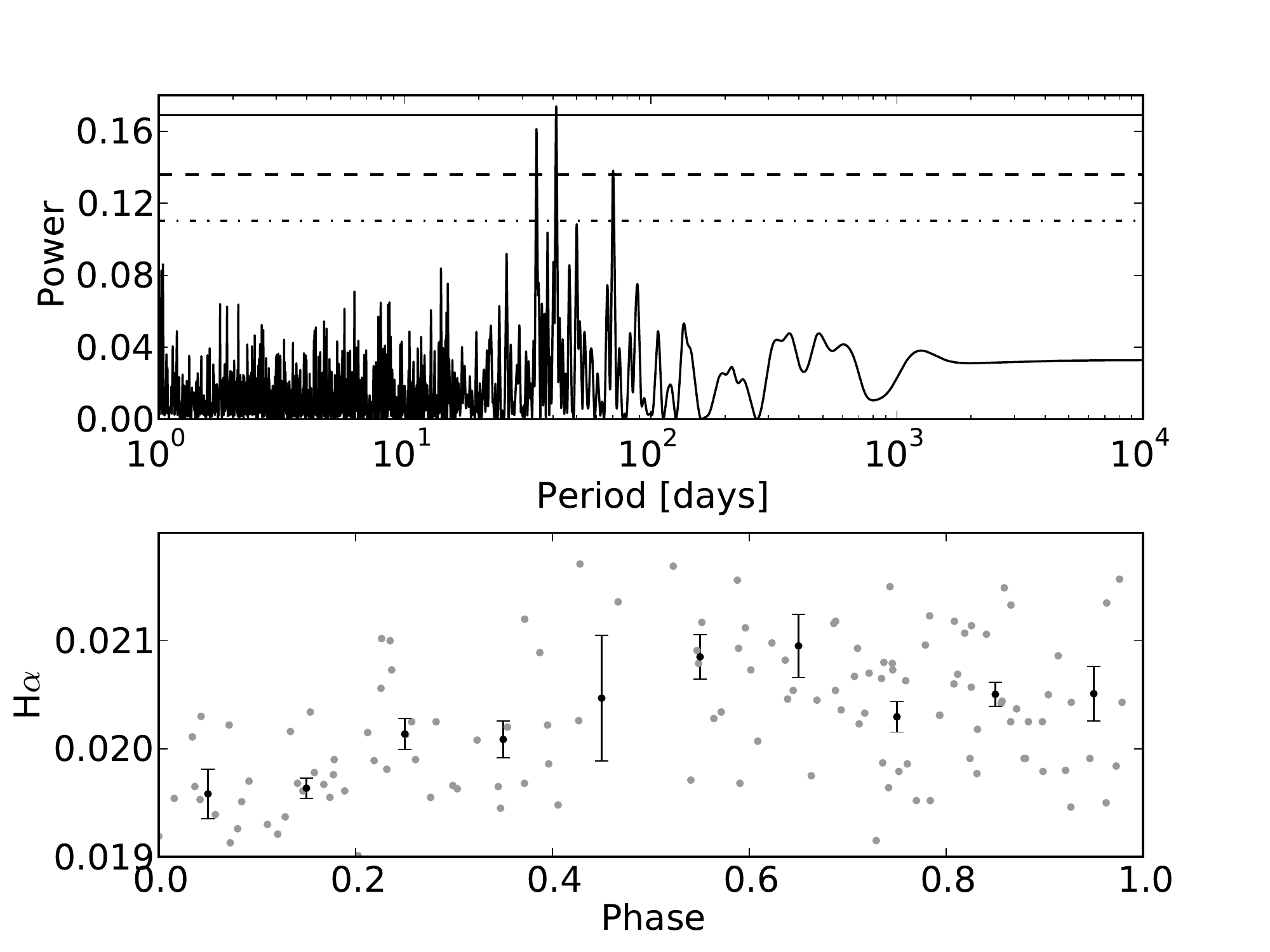}
\includegraphics[scale=0.45]{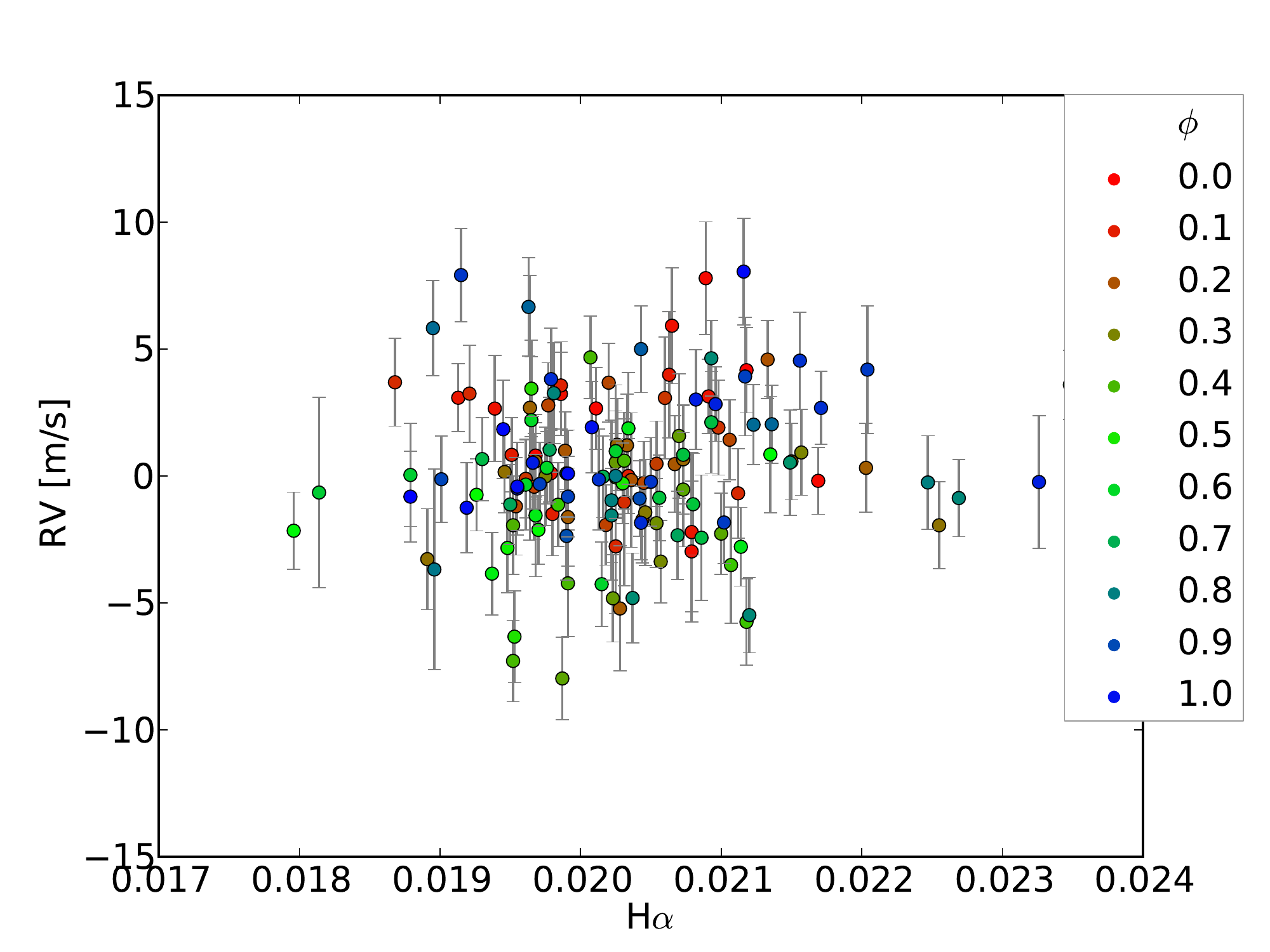}
\caption{\emph{Top:} Periodogram of the H$\alpha$ emission index of GJ~3293, 
with a peak above the 3$\sigma$ confidence level at 41 d and two peaks 
the above 2$\sigma$ confidence level at 34 and 70 d. \emph{Middle:} the 
H$\alpha$ index phased to P=41 d (grey dots); the black dots are binned by 0.1 in phase. \emph{Bottom:} RVs corrected for the 30.6 and 123.8~d signals 
against the $H\alpha$-index; the colours represent the phase of the 48~d
signal from table~\ref{tab:GJ3293_k3} (as represented in the middle panel of 
Fig.~\ref{fig:GJ3293_RV_phase})}
\label{fig:GJ3293_Halpha_gls}
\end{figure}

The periodograms of the bissector span, FWHM, contrast, and S-index 
show no dominant peaks, while that for H$\alpha$ shows one peak at
over 3$\sigma$ confidence at 41~d (Fig~\ref{fig:GJ3293_Halpha_gls} top); 
since the strength of the Ca~\textrm{\small II} emission in GJ~3293 is 
intermediate between those for Gl~176 ($P=39\;d$) and Gl~618A ($P=57\;d$) 
- Fig.~\ref{fig:activity_comparison}, this peak may
reflect the stellar rotation period. We see no correlation between 
any of the activity indicators and either the radial velocities or 
the residuals from subtracting the Keplerian orbits 
(Fig.~\ref{fig:GJ3293_Halpha_gls}, bottom, for $H\alpha$). 

To evaluate the stability of the 30.6, 48.1 and 123.8 d signals over 
time, we split the RVs into three groups of epochs: 2008-2009 
(34~measurements), 2010-2011 (52~measurements), and 2012-2013 (59~measurements).
We computed periodograms of the RVs and of their residuals after successively 
subtracting the stronger Keplerians. These seasonal periodograms consistently 
show strong evidence for the 30.6 and 123.8~d signals. The weaker 48.1~d
signal is detected in the 2008-2009 and 2012-2013 periodograms, but not in
the 2010-2011 epochs (bottom panels of Figs.~\ref{fig:GJ3293_RV_PeriodogramOBS},
~\ref{fig:GJ3293_RV_PeriodogramOMCk1} and~\ref{fig:GJ3293_RV_PeriodogramOMCk2}).
After investigating, we came to realize that the 2010-2011 measurements are 
strongly clustered at just two phases for a 48.1~d period and therefore
highly insensitive to that signal. There is consequently strong evidence 
for the stability and planetary nature of the 30.6 and 123.8~d signal,
but somewhat weaker evidence for the 48.1~d signal. The unfortunate phasing
of the 2010-2011 measurements and the weaker signal do not allow strong
limits against a possibly time varying amplitude. While the period of
that signal is moderately close to the 41~d possible stellar 
rotation and its true nature thus remains somewhat uncertain, the lack
of any significant correlation between radial velocity and the stellar 
activity indicators suggest that it is planetary.

\section{Radial velocities of GJ 3341}
\label{sec:GJ3341_analysis}

\begin{figure}[t]
\centering
\includegraphics[scale=0.45]{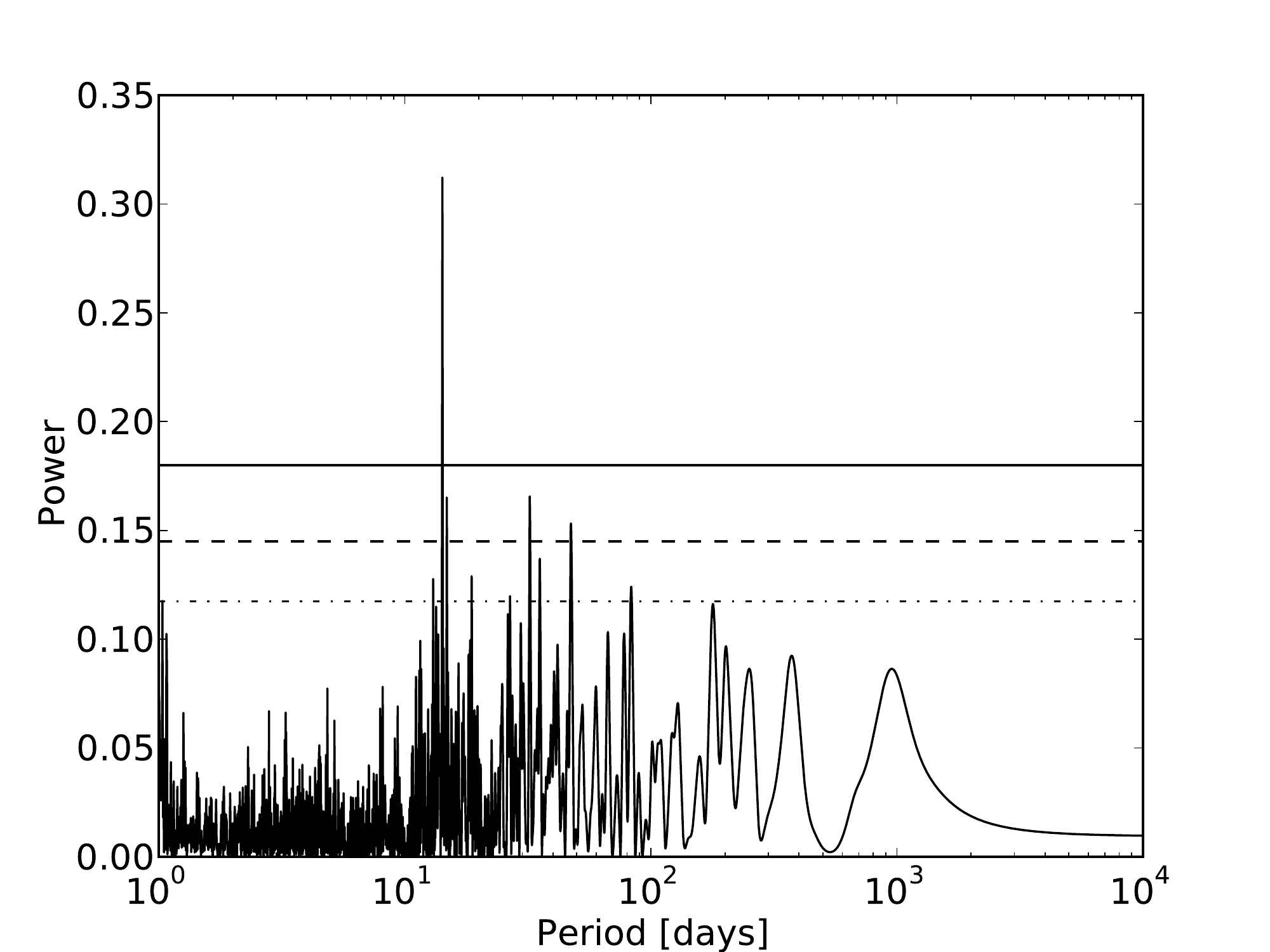}
\includegraphics[scale=0.45]{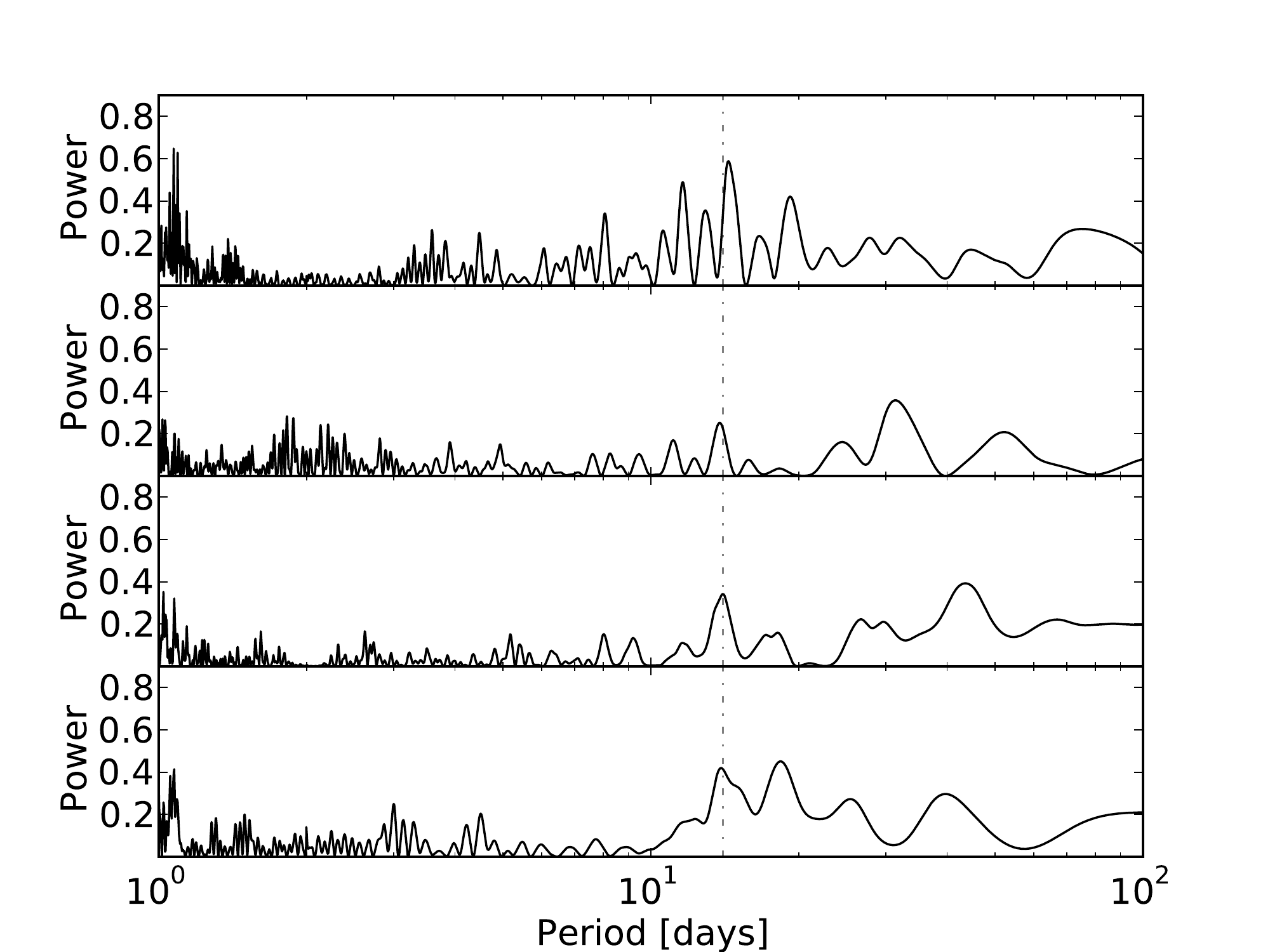}
\caption{\emph{Top panel:} Periodogram of the GJ~3341 radial velocities,
with a peak above the 3$\sigma$ confidence limit at 14~d 
(continuous black line). \emph{Bottom panel}: Periodograms 
for four subsets of epochs (BJD-2400000=54800-55000, 55400-55600, 
55800-56050, 56150-56300); the 14~d (vertical dashed dotted line) 
peak is always present.}
\label{fig:GJ3341_gls_obs}
\end{figure}

We obtained 135 RV measurements of GJ~3341, spanning 1456 d. Their 
dispersion is $\sigma_e=3.51ms^{-1}$, while the combined photon noise
and instrumental errors average to $\langle\sigma_i\rangle=1.89ms^{-1}$. 
An F-test and a $\chi^2$ comparison against a constant model yield
probabilities $P(F)$ and $P(\chi^2)<10^{-9}$ that the RV dispersion 
is explained by the RVs uncertainties. The periodogram 
(Fig.~\ref{fig:GJ3341_gls_obs}) shows a peak at 14.21 d with 
power of p=0.31. 1,000 iterations of bootstrap randomization 
produced no random data set with a power above 0.24, and the FAP 
for this peak is therefore well below $10^{-3}$. 
The \citet{1986ApJ...302..757H} recipe results in a FAP of $2.73\times10^{-8}$. 

\begin{figure}[t]
\centering
\includegraphics[scale=0.45]{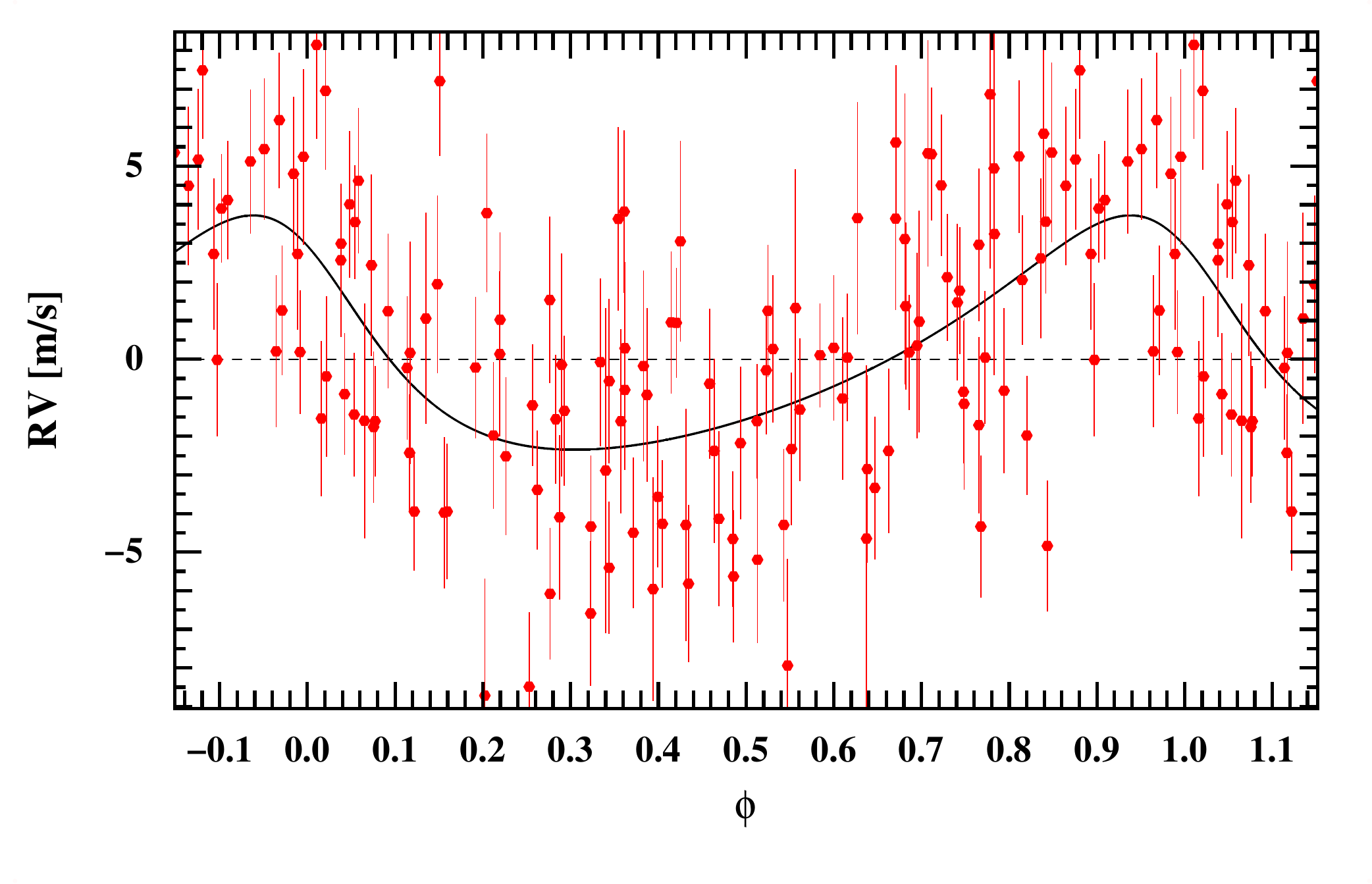}
\caption{ GJ~3341 radial velocities phased for a 14.2 d period.}
\label{fig:GJ3341_RV_phase}
\end{figure}

A Keplerian fit with \emph{yorbit} converges on an orbit with period 
$P=14.207\pm0.007\,d$, eccentricity $e=0.31\pm0.11$, and semi-amplitude 
$K_1=3.036\pm0.408$. Given the stellar mass of $M=0.47M_\odot$, the 
corresponding minimum planetary mass is $6.6\pm0.1M_\oplus$. 
Table~\ref{tab:GJ3341_k1} summarizes the solution parameters. This solution 
(Fig.~\ref{fig:GJ3341_RV_phase}) has a reduced chi-square of 
$\chi^2=2.28\pm0.19$ and a $\sigma_e=2.86ms^{-1}$ dispersion of the 
residuals. An F-test and a $\chi^2$ test for a constant model resulted 
in probabilities $P(F)=1.18\times 10^{-6}$ and $P(\chi^2)<10^{-9}$ that 
this dispersion is explained by photon noise combined with instrumental 
errors. The periodogram of the residuals shows a p=0.18 peak at 41~d,
above the p=0.16 level for a $1\%$ FAP and grazing the 3$\sigma$ confidence 
level. We could not reliably fit a  Keplerian to these residuals,
and stellar activity is therefore a more likely explanation for this 
additional RV variability.

\begin{table}[t]
\caption{One-keplerian fit for GJ~3341}
\label{tab:GJ3341_k1}
\centering
\begin{tabular}{l l}
\hline
\noalign{\smallskip}
& GJ~3341b \\
\noalign{\smallskip}
\hline\hline
\noalign{\smallskip}
P [d] & $14.207\pm0.007$\\
$T_0$ [JD-2400000] & $55622.7\pm0.8$\\
$\omega$ [deg] & $42.32\pm4.3$\\
e & $0.31\pm0.11$\\
$K_1$ [$ms^{-1}$] & $3.04\pm0.41$\\
$m\, sin(i)$ [$M_\oplus$] & $6.6\pm0.1$\\
a [AU] & 0.089\\
\noalign{\smallskip}
\hline
\noalign{\smallskip}
$\gamma \; [kms^{-1}]$ & $47.803\pm0.003$ \\
$N_{meas}$ & 135\\
Span [d] & 1455.9\\
$\langle\sigma_i\rangle\,[ms^{-1}]$ & 1.89\\
$\sigma_e \, [ms^{-1}]$ & 2.86\\
$\chi_\nu^2$ & 2.28\\
\noalign{\smallskip}
\hline
\end{tabular}
\end{table}

\subsection{Stellar activity}

\begin{figure}[t]
\centering
\includegraphics[scale=0.45]{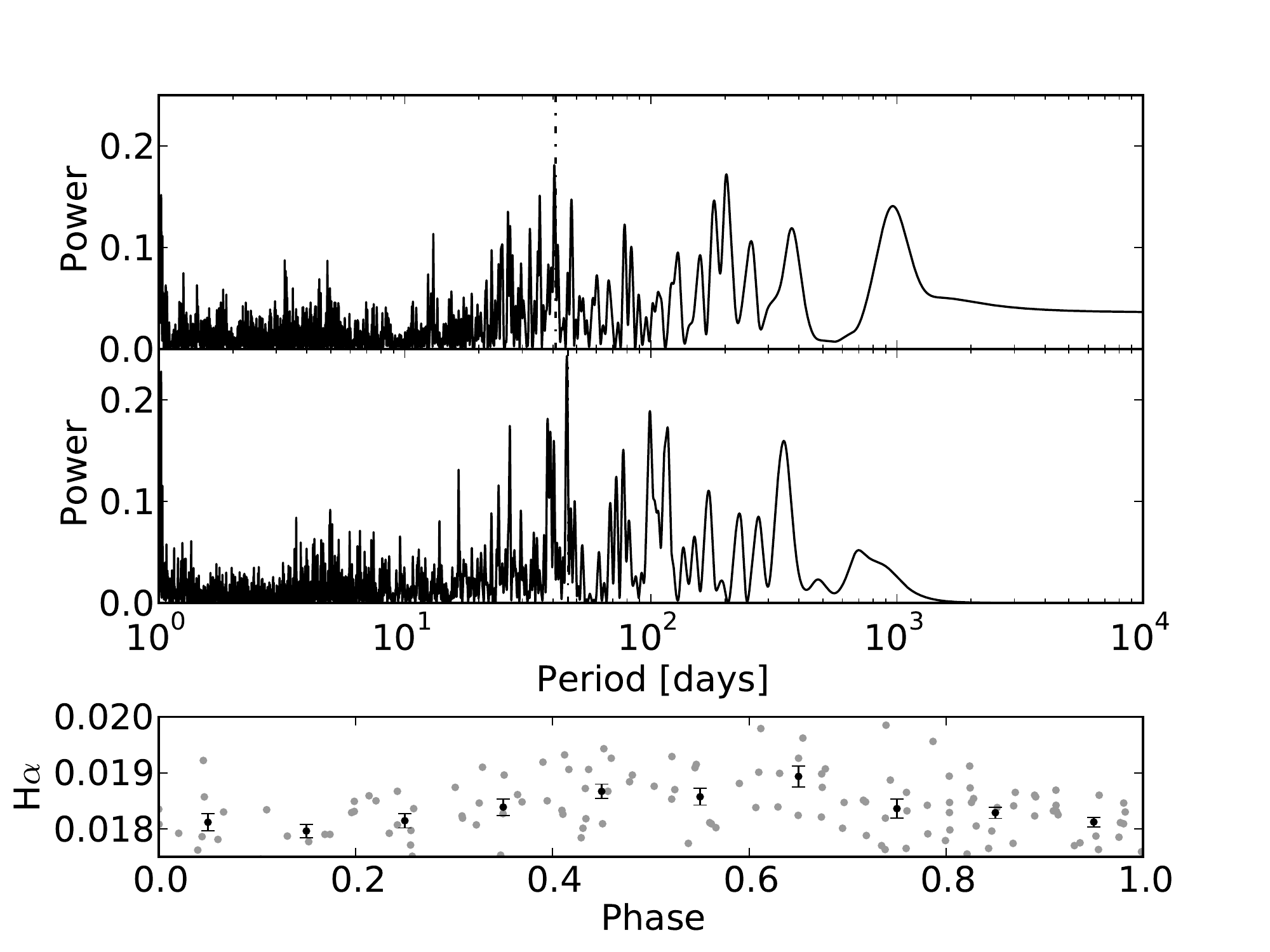}
\includegraphics[scale=0.45]{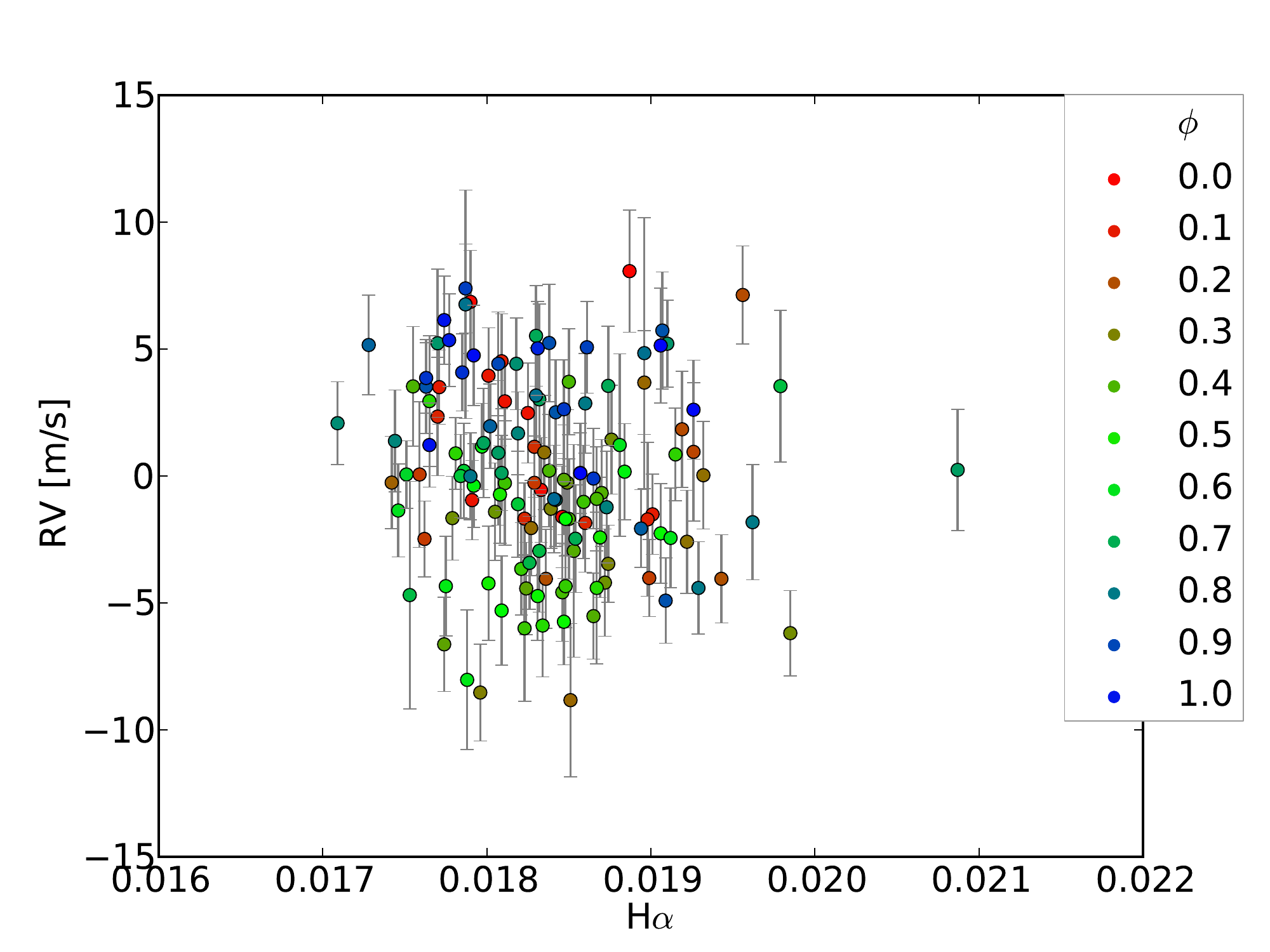}
\caption{\emph{Top:} Periodogram of the H$\alpha$ emission index for GJ~3341,
with a peak above the 3$\sigma$ confidence level (continuous line) at 46 d. 
\emph{Middle:} The H$\alpha$ emission index phased to 46~d, which may
represent the rotation period. 
\emph{Bottom:} RV against $H\alpha$ index, with colours represents the phase 
for the 14~d signal as in figure~\ref{fig:GJ3341_RV_phase}, 
$T_0$ from table~\ref{tab:GJ3341_k1}. }
\label{fig:GJ3341_Halpha_gls_phase}
\end{figure}

The periodogram of the $H\alpha$ (Fig.~\ref{fig:GJ3341_Halpha_gls_phase}) and 
$S$ indices, contrast, bissector-span, and FWHM of the CCF show no evidence 
of stellar activity which could explain the RVs variations, and  nor do 
plots of the RV as a function of these parameters. Subtracting a long term 
trend visible in $H\alpha$ index, however, increases the power in a 
pre-existing 46~d peak of its periodogram 
(Fig.~\ref{fig:GJ3341_Halpha_gls_phase}) to 0.24, above the 3$\sigma$ 
confidence level. Phasing the H$\alpha$ index with this period produces 
relatively smooth and approximately sinusoidal variations, compatible 
with the signature of stellar rotation. This period is somewhat shorter, 
than expected from the relatively weak Ca~\textrm{\small II} emission 
of GJ~3341, which is intermediate in strength between those of Gl~618A 
(P=57 d) and Gl~581 (P=130 d) (Fig.~\ref{fig:activity_comparison}), but 
probably within the dispersion of the period-activity relation. The 
period closely matches that found in the RV residuals, reinforcing
activity as an explanation for those. 

To evaluate the stability of the 14~d signal, we split the RVs into 
four seasons (BJD-2400000=54800-55000, 55400-55600, 55800-56050, 
56150-56300; 25, 32, 43, 32 measurements per epoch) and computed 
periodograms for each. The 14~d is consistently  present in 
every periodogram.

\section{Radial velocities of GJ 3543}
\label{sec:GJ3543_analysis}

\begin{figure}[t]
\centering
\includegraphics[scale=0.45]{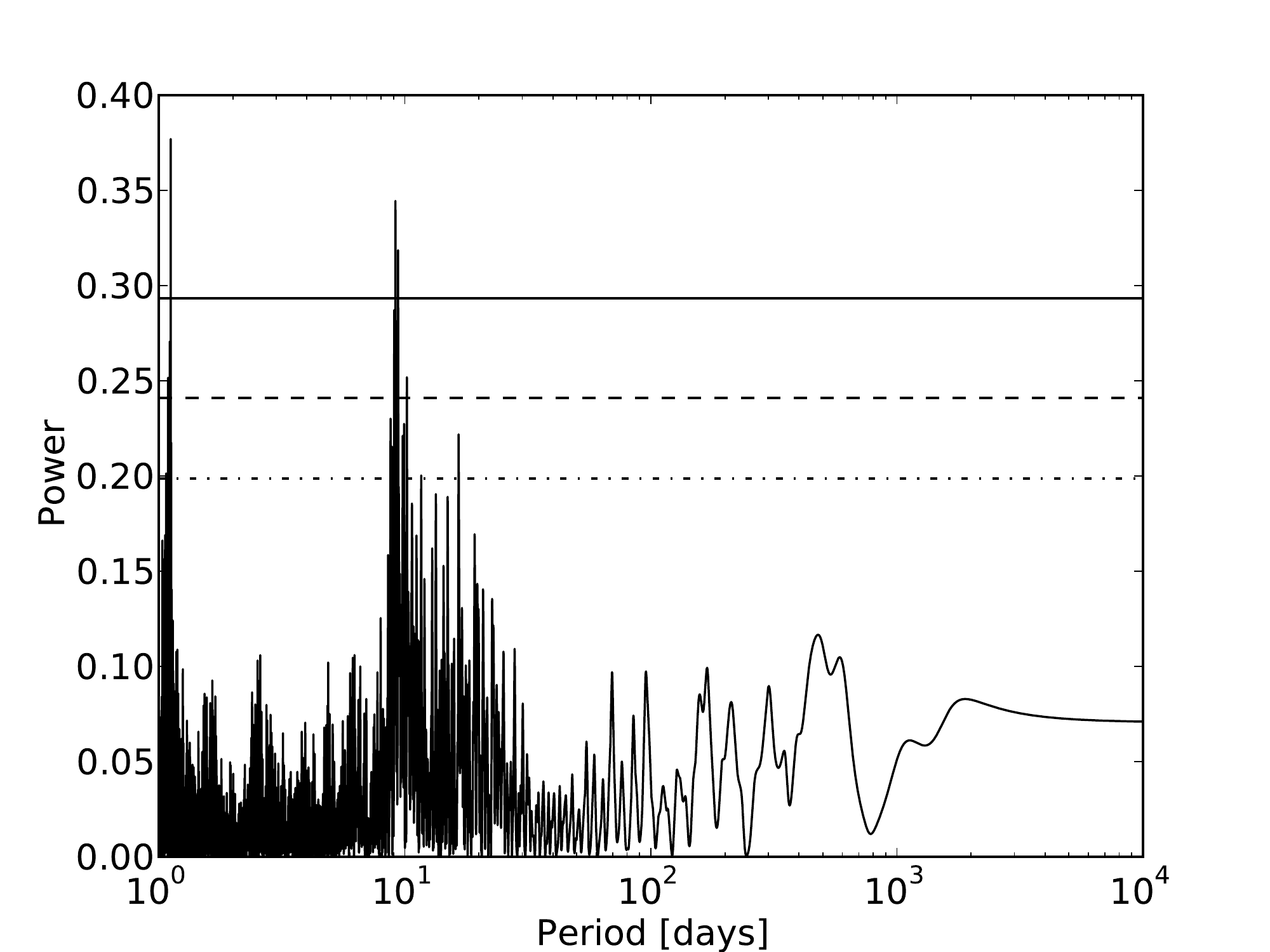}
\includegraphics[scale=0.45]{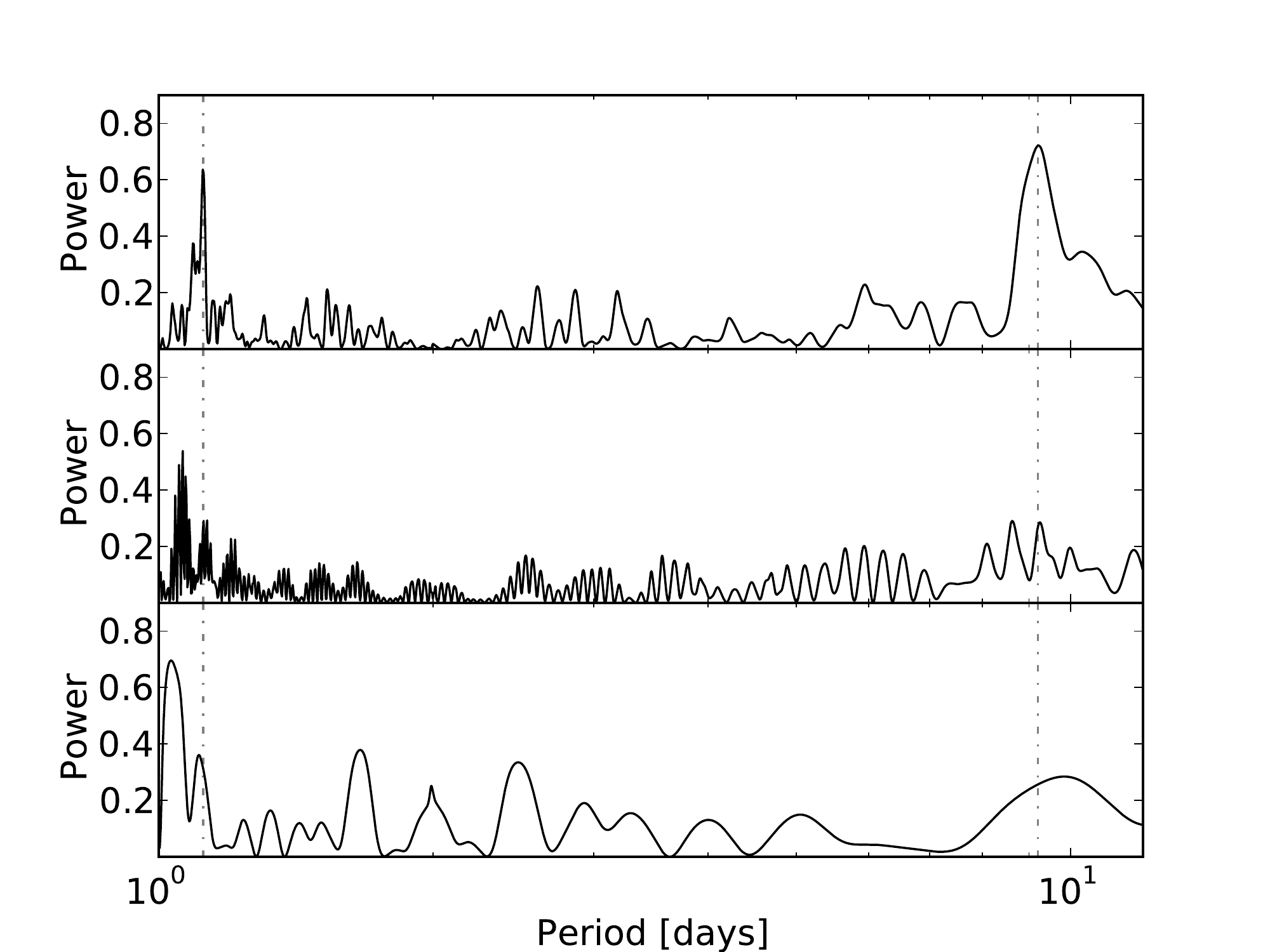}
\caption{\emph{Top panel:} Periodogram of the radial velocities of 
GJ~3543. The horizontal lines represent the same false alarm levels as in 
Fig.~\ref{fig:GJ3293_RV_PeriodogramOBS}. The two peaks at 1.1 and 9.2~d
have less than $0.3\%$ FAP. 
\emph{Bottom panel:} Periodograms for three independent subsets of
observation epochs (BJD-2400000=55500-55750, first row; 
55850-56050, second row; 56340-56380, third row). The 1.1 and 9.2~d 
peaks (marked by vertical dashed-dotted lines) are both unstable over 
time.}
\label{fig:GJ3543_RV_periodogram}
\end{figure}

We obtained 80~RV measurements of GJ~3543 spanning 1919~d, with a 
dispersion $\sigma_e=3.02\,ms^{-1}$ compared to an average photon noise
combined with instrumental error of $\langle\sigma_i\rangle=1.21ms^{-1}$. 
An F-test and a chi-square test for a constant model find a 
$P<10^{-9}$ probability that these known measurement errors explain 
the dispersion. The periodogram (Fig.~\ref{fig:GJ3543_RV_periodogram}) 
of the GJ~3543 RVs exhibits two strong peaks with powers of 0.37 and 0.34
at 1.1 and 9.2 d. Both peaks are well above the p=0.29 for power at 0.3$\%$ FAP confidence level.

\begin{figure}[t]
\centering
\includegraphics[scale=0.45]{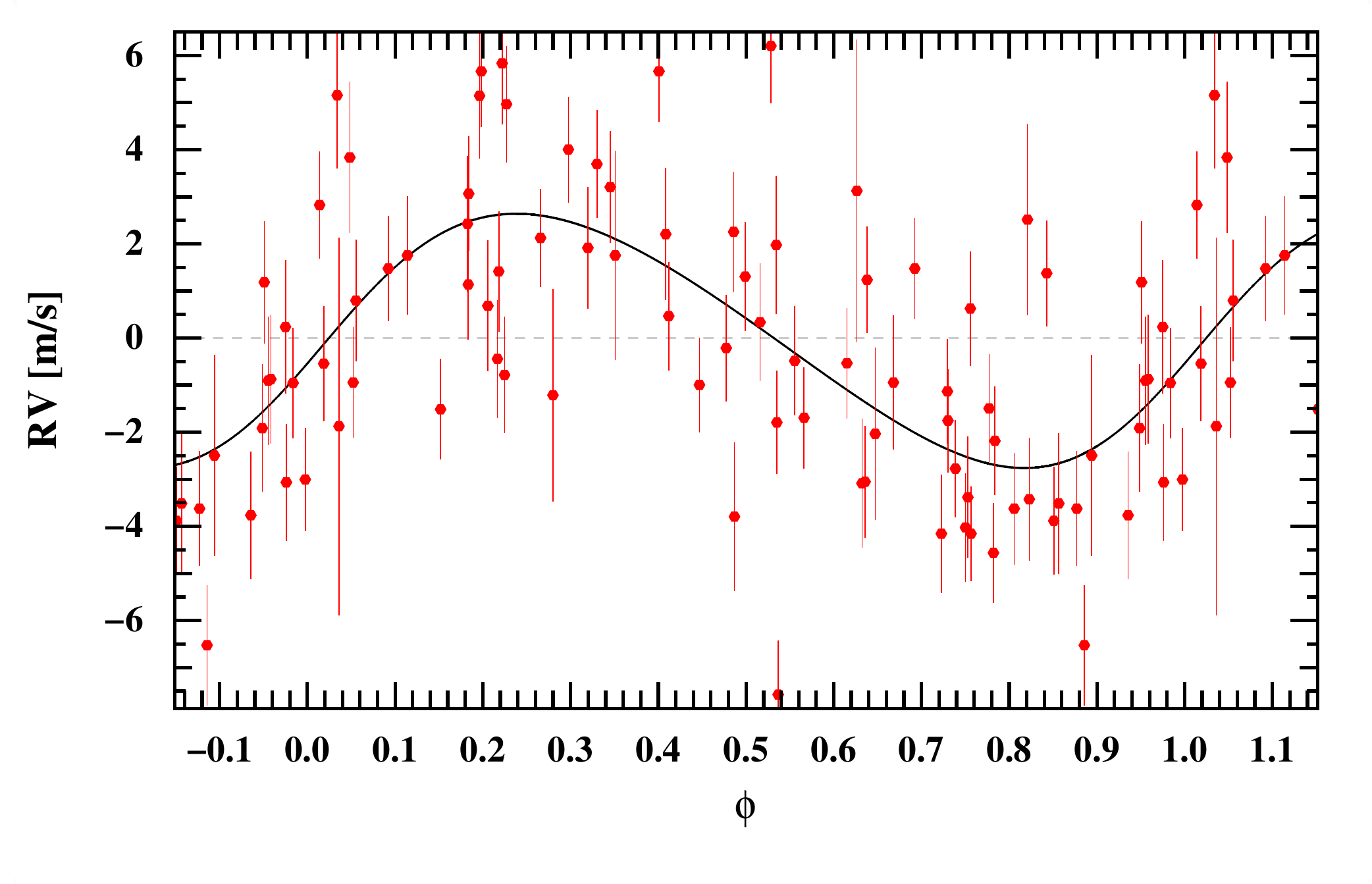}
\includegraphics[scale=0.45]{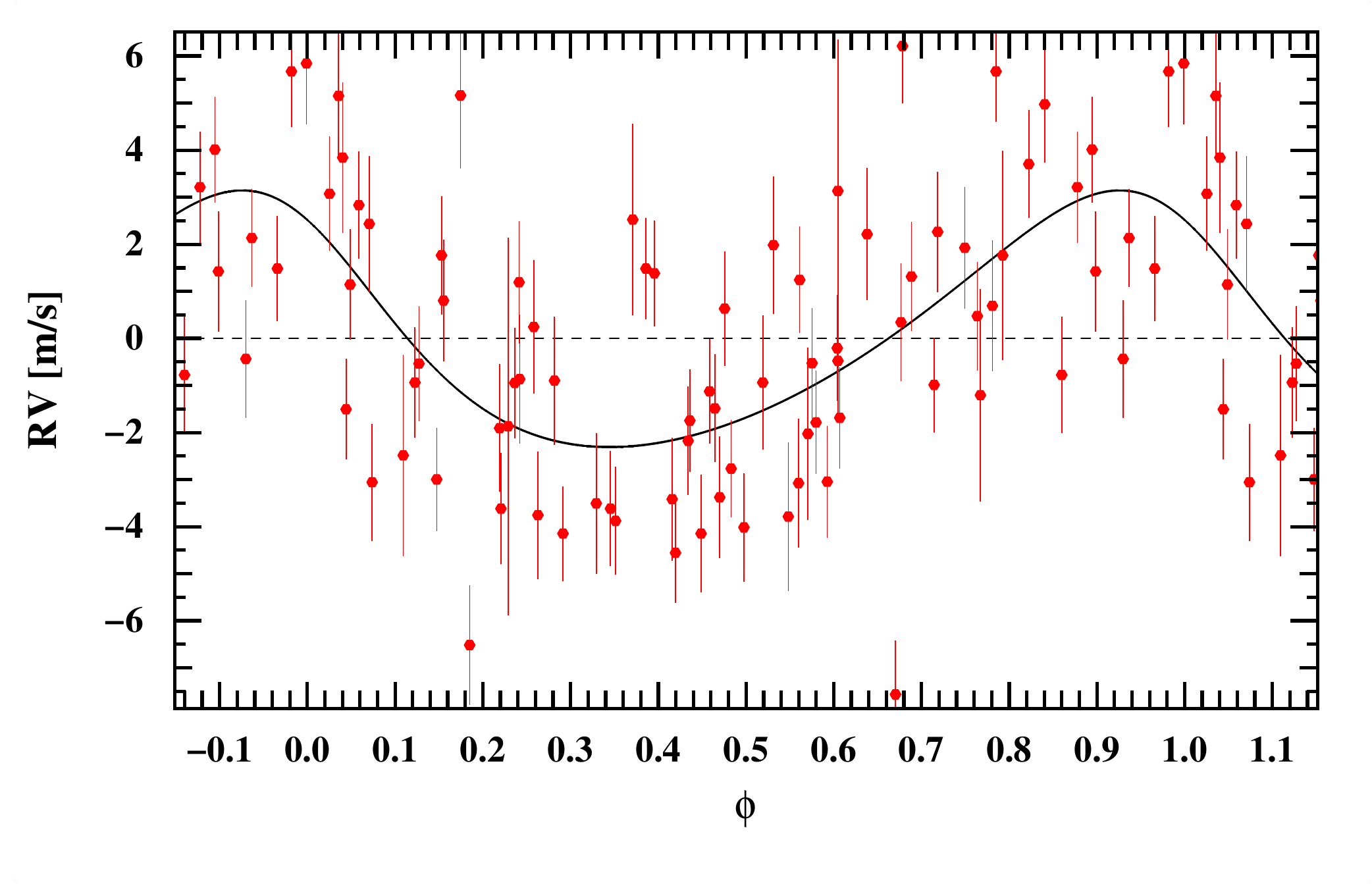}
\caption{Radials velocities for GJ~3543 RVs phased to the 1.1 d (top) 
and 9.2 d (bottom) period.}
\label{fig:GJ3543_RV_phase}
\end{figure}

Our first \emph{yorbit} one-Keplerian fit to the RVs converged to an
orbit with period $P=1.11913\pm0.00006$, eccentricity $e=0.13\pm0.16$, 
and semi-amplitude $K_1=2.70\pm0.38$ ($2.6\pm0.4M_\oplus$). This solution 
decreases the dispersion to $\sigma_e=2.32\,ms^{-1}$ and the reduced 
chi-square of the residuals to $\chi^2=3.80\pm0.32$. The 9.2~d signal
disappears in the periodogram of the residuals, demonstrating that
the 1.1 and 9.2~d peaks are aliases of each other. If we introduce a 
prior that mildly favors a longer period, the fit instead converges 
to an orbit with period $P=9.161\pm0.004$, eccentricity $e=0.20\pm0.15$, 
and semi-amplitude $K_1=2.73\pm0.44$ ($5.1\pm0.9M_\oplus$), and the 
1.1~d signal disappears in the periodogram of the residuals. Their 
dispersion is $\sigma_e=2.42\,ms^{-1}$ and the reduced chi-square is 
$\chi^2=4.14\pm0.33$. In either case the strongest peak in the periodogram
of the residuals occurs at 23~d and has a power p=0.25 which corresponds 
to a $\sim$2.5$\%$ FAP. The daily sampling of the observations and the 
periodogram analysis both suggest that the 1.1 and 9.2~d signals are 
aliases of one other (${1}/{1.119} + {1}/{9.161}={1}/{1.003}$). 
The residuals of the two fits don't differ enough to ascertain which
represents the true signal, and Fig.~\ref{fig:GJ3543_RV_phase} therefore
plots both solutions.

\begin{table}[t]
\caption{Fit for one keplerian of GJ~3543}
\label{tab:GJ3543_k1}
\centering
\begin{tabular}{l l l}
\hline
\noalign{\smallskip}
& GJ~3543(b) & GJ~3543(b) \\
\noalign{\smallskip}
\hline\hline
\noalign{\smallskip}
P [d]                 & $1.11913\pm0.00006$ & $9.161\pm0.004$ \\
$T_0$ [JD-2400000]       & $55739.52\pm0.18$   & $55745.6\pm1.0$ \\
$\omega$ [deg]           & $-100\pm60$         & $39.7\pm43.3$    \\
e                        & $0.13\pm0.16$       & $0.20\pm0.15$    \\
$K_1$ [$ms^{-1}$]         & $2.70\pm0.38$       & $2.73\pm0.44$     \\
$m\, sin(i)$ [$M_\oplus$] & $2.6\pm0.4$         & $5.1\pm0.9$         \\
a [AU]                   &0 .0162              & 0.0657           \\
\noalign{\smallskip}
\hline
\noalign{\smallskip}
$\gamma \; [kms^{-1}]$ & $15.0927\pm0.0004$ & $15.0932\pm0.0005$ \\
$N_{meas}$                & 80             &  80                   \\
Span [d]             & 1918.8          & 1918.8                \\
$\langle\sigma_i\rangle\,[ms^{-1}]$ & 1.21 & 1.21\\
$\sigma_e \, [ms^{-1}]$             & 2.32 & 2.42\\
$\chi_\nu^2$                        & 3.80 & 4.14\\
\noalign{\smallskip}
\hline
\end{tabular}
\end{table}

\subsection{Stellar activity}

\begin{figure}[t]
\centering
\includegraphics[scale=0.45]{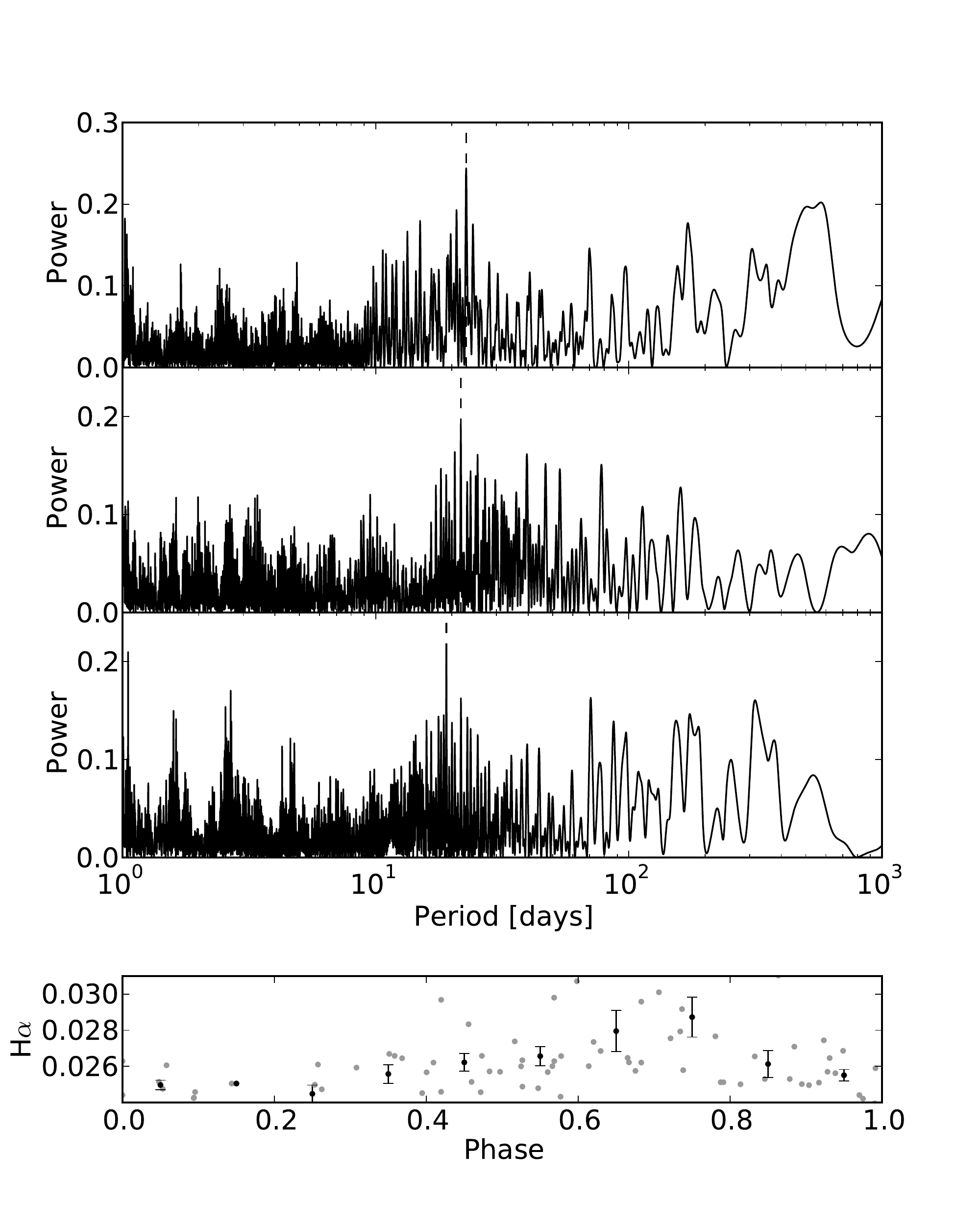}
\includegraphics[scale=0.45]{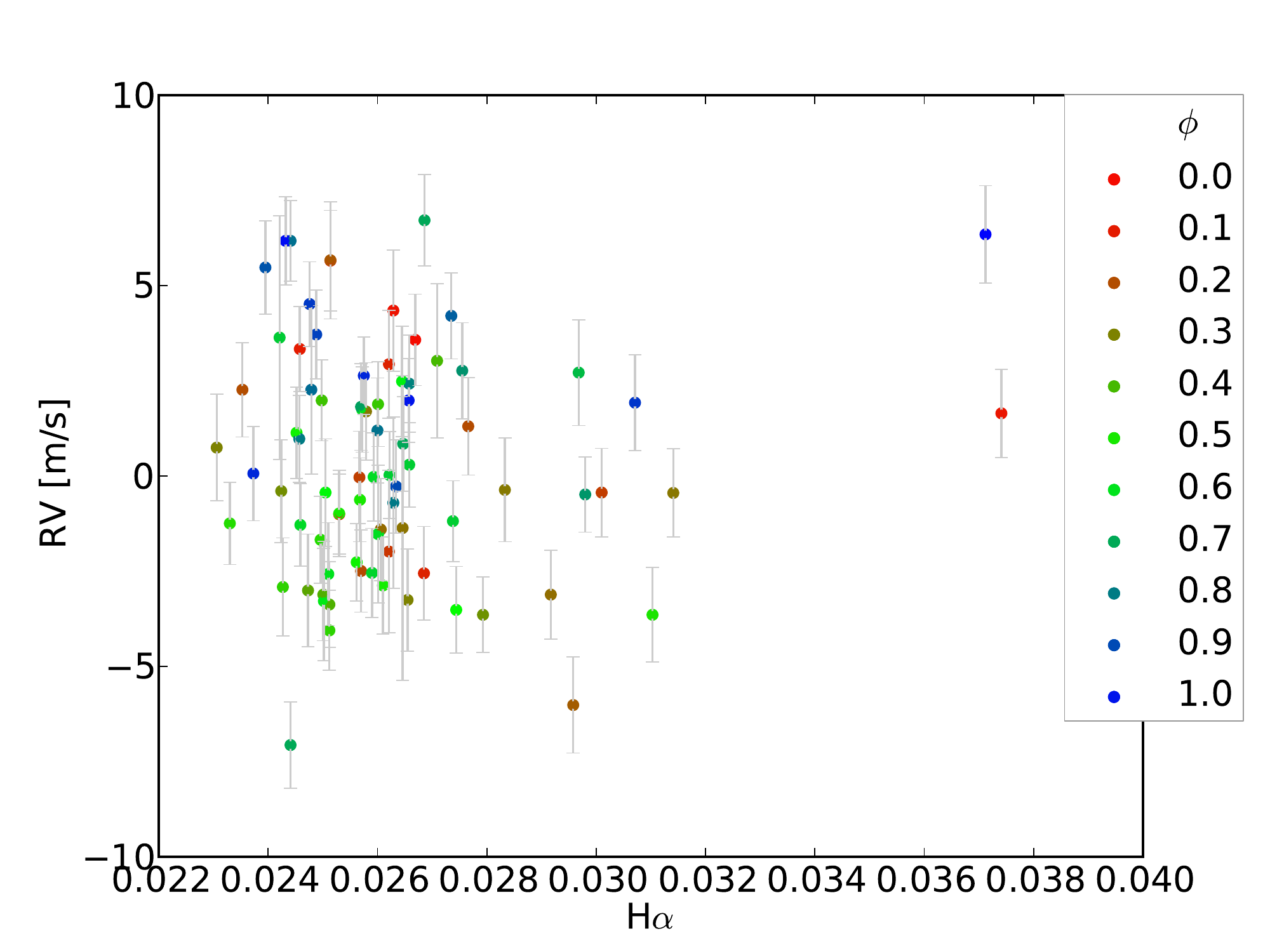}
\caption{\emph{Top panel:} \emph{First row}: Periodogram of the residuals
of the GJ~3543 radial velocities after subtracting the 1.1~d Keplerian 
orbit. The power excess at P=23 $d$ has a 2.5$\%$ FAP. 
\emph{Second row}: Periodogram of the S-index of  GJ~3543. The significance
of the 22~d peak is just under $1\sigma$. 
\emph{Third row}: Periodogram of the H$\alpha$ index of  GJ~3543. The 
false alarm probability of the 19~d peak is 14\%. 
\emph{Middle panel}: The H$\alpha$ index phased to the 19~d period. 
\emph{Bottom panel:} RVs against the $H\alpha$-index, the colours 
represents phase for the 9.2~d period, as represented in 
Fig.~\ref{fig:GJ3543_RV_phase} (bottom), $T_0$ from table~\ref{tab:GJ3543_k1}.}
\label{fig:GJ3543_Halpha}
\end{figure}

The power in the strongest peak in the periodogram of the S-index, 
(Fig.~\ref{fig:GJ3543_Halpha}, second row), at 22~d, is p=0.193
and just below the p=0.198 needed for the $1\sigma$ confidence level.
The strongest peak in the periodogram of the H$\alpha$ index 
(Fig.~\ref{fig:GJ3543_Halpha}, third row), at 19~d, is above 
the $1\sigma$ confidence level but still has a 14$\%$ FAP. Either
period would be consistent with the strength of the Ca~\textrm{\small II} 
emission line (Fig.~\ref{fig:activity_comparison}), which suggest a 
stellar rotation period shorter than 35~d. While both activity 
signals have low significance, one can note that the P=9.2~d 
radial velocity period is close the the first harmonic of either
19 or 22~d, and that the tentative 23 d peak in the periodogram 
of the RV residuals (Fig.~\ref{fig:GJ3543_Halpha}, first row) is also 
close to both. We evaluated the stability of the 1.1 or 9.2~d signal
by computing periodograms for three disjoint seasons, 
BJD-2400000=55500-55750, 55850-56050, and 56340-56380, which contain 
25, 30, and 14 measurements. The two aliased signals are present in 
the first season only, and absent in the second and third seasons 
(Fig.~\ref{fig:GJ3543_RV_periodogram}, bottom panel). The seasonal
datasets have too few measurements for a similar exercise for the 
tentative 23~d peak in the periodogram of the residuals. 
Our best guess is that stellar activity is responsible for
the RVs variation, although we see no correlation between the
variations of the RV and of the $S$ or $H\alpha$ indices. More 
data will be needed to ascertain the source of the RV dispersion.

The radial velocity signal at half the stellar rotation period
found here for GJ~3543 has an analog in the recent reanalysis by 
\citet{2014Sci...345..440R} of the \citet{2011arXiv1109.2505F} GJ~581 
data. This analogy provides an opportunity to summarize here the views
of our team on the physical reality of the up to 6 planets that have 
been claimed to to orbit GJ~581, with heated controversies on the
statistical significance of the weaker signals.

Our group announced the discoveries of 'b' in 2005, followed by 'c' and 
'd' in 2007 and then 'e' in 2009 \citep{2005A&A...443L..15B,
2007A&A...469L..43U,2009A&A...507..487M}, after considering
both planetary and activity models in the interpretation of the 
observed periodic signals. The estimated rotational period of GJ~581 
was much longer than the putative orbital periods of b, c and e, 
which consequently were immediately accepted as planets. The 
interpretation of the 'd' signal was less straightforward, because 
it occured at a plausible rotational periods for GJ~581. We discarded
that explanation at the time, on the grounds that the Doppler 
variations, if caused by a spot on the rotating star, would have
come together with larger photometric variations than observed
\citep[e.g.][]{2007A&A...474..293B}. This made the planet 
the most likely interpretation, at that time.

\citet{2010ApJ...723..954V} then combined the 2004 to 2008 
HARPS data with new HIRES data, with most of the statistical weight on the
HARPS side, to announce two additional planets in the system, f 
and g. We were monitoring GJ~581 very intensively, and we quickly 
reacted to \citet{2010ApJ...723..954V}'s announcement by showing 
that our new HARPS data were incompatible with these additional
two planets \citep{2011arXiv1109.2505F}. In that manuscript we 
relied on Vogt et al.'s measurement of a 90~day rotational period 
for GJ~581 to conclude that GJ~581d was a bonafide planet, since
its period was comfortably away from any harmonic of the presumed
rotational period.

Strong doubts, on different grounds, on the reality of GJ~581 f and g 
were also expressed by others \citep{2011A&A...528L...5T,
2011MNRAS.415.2523G,2013MNRAS.429.2052B}. Baluev et al. additionally 
questioned whether GJ~581~d exists, finding that accounting 
for the correlated noise in the radial velocity measurements of
GJ~581 decreased the significance of 'd' to $\sim$1.5 $\sigma$. 
Robertson et al. more recently identified the astrophysical source 
of that correlated noise, showing that GJ~581 obeys a more complex
RV-activity relation than previously thought. Instead of star spots, 
they invoke convection inhibition within active regions that locally 
changes the balance of ascending vs. descending material. Such active 
regions move as the star rotates and induce apparent Doppler shifts,
but do not necessarily induce brightness variations. Robertson et al.
additionnally find the true rotation period of GJ~581 to be 130~day, 
quite different from that announced by Vogt et al. and twice the 
period of 'd'. These findings together mean that the 65~days radial
velocity signal is most probably due to 2 longitudinally opposed 
active regions, and show that extra caution is warranted when RV 
periodicity are found near a harmonic of the rotation period. This 
occurs here for GJ~3543, and might also be the case for the GJ~667C 
system \citep{2013A&A...553A...8D, 
2013A&A...556A.126A, 2014MNRAS.437.3540F}, though its rotation
period remains slightly uncertain.

\section{Summary and conclusions}
\label{sec:Conclusion}
We analysed observations of three early-M dwarfs with the HARPS spectrograph 
mounted on the $3.6m$ telescope at La Silla observatory (ESO). We identify
a planetary system orbiting GJ~3293, composed of two neptunes 
with periods near the 4:1 resonance ($30.6\pm0.02$ and $123.98\pm0.38$ d), 
and more tentatively a super-Earth with an orbital period of 
$48.14\pm0.12$ d. Although the RV variations appear uncorrelated with 
any stellar activity indicator, the orbital period of the least massive 
planet candidate remains moderately close to the plausible stellar 
rotation period. This signal is present and stable for the 2008-2009 
and 2012-2013 subsets of the data, while the 2010-2011 subset has 
inadequate sampling to probe a 48~d period. More data will be needed 
to fully confirm this planet candidate. With a  $0.194 \;AU$  semi-major 
axis it orbits in the habitable zone of  GJ~3293, and with a minimum mass 
of $7.9\pm1.4M_\oplus$ it could be rocky. The hierarchical structure of 
the system warrants a dynamical analysis.
 
GJ~3341 is orbited by a uper-Earth ($msin(i)\sim6.6M_\oplus$),
which its $14.207\pm0.007$~d period places in the inner habitable zone 
of its host star. 

The periodogram of the radial velocities of GJ~3543 is dominated by two
mutually aliased peaks at 1.1 and 9.2~d, but those  are only present 
in a subset of the epochs. The periodograms of the stellar activity indices 
suggest a stellar rotation period of about 20 d, or approximately
twice the 9.2~d period, which further reinforces the presumption 
that stellar activity is responsible for the unstable radial velocity 
signal - see \citet{2011A&A...528A...4B}. 

GJ~3293 and GJ~3341 have approximately solar-metallicity, consistently
with the observation that the frequency of super-Earth and neptune 
planets seems uncorrelated with stellar metallicity 
\citep{2011arXiv1109.2497M, 2011A&A...533A.141S, 2013A&A...551A..36N}. 
As the sample of well characterized planetary systems increases and stellar 
properties are more accurately known, we will refine the statistical
relations between the presence of planets and the stellar properties of 
their hosts, which will help constrain planet formation and evolution models.

\begin{acknowledgements}
N. A. acknowledges support from CONICYT Becas-Chile 72120460. This publication makes use of data products from the Two Micron All Sky Survey, which is a joint project of the University of Massachusetts and the Infrared Processing and Analysis Center/California Institute of Technology, funded by the National Aeronautics and Space Administration and the National Science Foundation. X. B., X. D., and T. F. acknowledge the support of the French Agence Nationale de la Recherche (ANR), under the program ANR-12-BS05-0012 Exo-atmos. X.B. acknowledges funding from the European Research Council under the ERC Grant Agreement n. 337591-ExTrA. NCS acknowledges the support from the European Research Council/European Community under the FP7 through Starting Grant agreement number 239953. NCS further acknowledges the support from Funda\c{c}\~ao para a Ci\^encia e a Tecnologia (FCT, Portugal) through FEDER funds in program COMPETE, as well as through national funds, in the form of grants reference RECI/FIS-AST/0176/2012 (FCOMP-01-0124-FEDER-027493), and RECI/FIS-AST/0163/2012 (FCOMP-01-0124-FEDER-027492), and through the Investigador FCT contract reference IF/00169/2012 and POPH/FSE (EC) by FEDER funding through the program "Programa Operacional de Factores de Competitividade - COMPETE. 
\end{acknowledgements}

\bibliographystyle{aa}
\bibliography{GJ3293_GJ3341_GJ3543}

\begin{appendix}

\section{RVs}
\label{sec:RV_appendix}
The RVs in the barycentric frame and corrected for secular acceleration. 
The RVs were extracted through chi$^2$ matching to a high signal-to-noise
ratio templing. The errors combine the estimated photon noise with the 
an estimate of the residual instrumental error ($0.60ms^{-1}$). We also
tabulate the FWHM and bisector span (BIS) of the cross-correlation 
function, as well as the $H\alpha$ and Ca~\textrm{\small II} 
\emph{Mount Wilson} S activity indices.

\longtab{1}{
\begin{longtable}{lllllllll}
\caption{GJ 3293 RVs, their uncertainty and activity indicators.}\\
\hline\hline
BJD - 2400000 & RV &  $\sigma_{RV}$ & FWHM& Contrast  & BIS  & S-index & $H\alpha$ \\
 & $[kms^{-1}]$ & $[kms^{-1}]$ & [$kms^{-1}$] & & [$kms^{-1}$] &&  \\
\hline
\endfirsthead
\caption{continued.}\\
\hline\hline
BJD - 2400000 & RV &  $\sigma_{RV}$ & FWHM & Contrast & BIS & S-index & $H\alpha$\\
 & $[kms^{-1}]$ & $[kms^{-1}]$   $[kms^{-1}]$ & & [$kms^{-1}$] & [$kms^{-1}$] & &\\

\hline
\endhead
\hline
\endfoot
54805.68424	& 13.28536	& 0.00375	& 3.60254&	25.79888& 	-17.14562		& 1.42156	&	0.01814 \\ 
54825.64027	& 13.30846	& 0.00132	& 3.60782&	25.14589& 	-5.88699		& 1.46394	&	0.02169 \\ 
54826.63445	& 13.31021	& 0.00146	& 3.60512&	25.07722& 	-3.29884		& 1.33697	&	0.02091 \\ 
54827.64845	& 13.30529	& 0.00148	& 3.60866&	24.69961& 	-11.02001		& 1.16751	&	0.02034 \\ 
54828.66060	& 13.30270	& 0.00176	& 3.59488&	24.71040& 	-3.73801		& 1.42195	&	0.02112 \\ 
54829.76990	& 13.30312	& 0.00187	& 3.59572&	25.10511& 	-6.39420		& 1.17184	&	0.02098 \\ 
54830.66322	& 13.29993	& 0.00168	& 3.60334&	24.87551& 	-5.67922		& 1.20943	&	0.02054 \\
54831.65871	& 13.29726	& 0.00163	& 3.61586&	24.87951& 	-11.28642		& 1.19192	&	0.02045 \\
54832.67287	& 13.29552	& 0.00265	& 3.60512&	25.07722& 	-3.29884		& 1.28864	&	0.02036 \\
54833.66016	& 13.29525	& 0.00201	& 3.59258&	24.73141& 	-8.97254		& 1.16719	&	0.02033 \\
54834.71626	& 13.29312	& 0.00151	& 3.60281&	25.05025& 	-12.85689		& 1.28836	&	0.02150 \\
54848.62098	& 13.30394	& 0.00142	& 3.60453&	24.78696& 	-12.39565		& 1.04256	&	0.01926 \\
54850.61453	& 13.30143	& 0.00163	& 3.60655&	24.93891& 	-3.71860		& 1.00371	&	0.01937 \\
54852.63392	& 13.30466	& 0.00142	& 3.59689&	25.03757& 	-8.62567		& 0.99302	&	0.01976 \\
54854.62660	& 13.30131	& 0.00169	& 3.58326&	24.82394& 	-7.40420		& 1.16644	&	0.02056 \\
54879.59166	& 13.30268	& 0.00168	& 3.59159&	25.00795& 	-14.86627		& 1.27681	&	0.01977 \\
54881.62441	& 13.29896	& 0.00193	& 3.62406&	24.69537& 	-12.85689		& 1.28265	&	0.01991 \\
54883.52494	& 13.29997	& 0.00160	& 3.60304&	24.86785& 	-8.97254		& 1.11932	&	0.01946 \\
54885.56408	& 13.29871	& 0.00169	& 3.61087&	24.90341& 	-7.36429		& 1.31220	&	0.02157 \\
55045.92231	& 13.28938	& 0.00261	& 3.59152&	24.95034& 	-5.09714		& 0.81714	&	0.02025 \\
55047.91349	& 13.28374	& 0.00247	& 3.59427&	24.86074& 	-5.09714		& 1.37098	&	0.02086 \\
55049.92289	& 13.28389	& 0.00158	& 3.60506&	24.76291& 	-7.40420		& 1.17482	&	0.01950 \\
55052.90367	& 13.28542	& 0.00152	& 3.59590&	24.69413& 	-19.69584		& 1.18646	&	0.02269 \\
55121.83805	& 13.29732	& 0.00190	& 3.61108&	25.01686& 	-12.34296		& 1.21001	&	0.02067 \\
55126.73175	& 13.29736	& 0.00163	& 3.60254&	25.79888& 	-17.14562		& 1.42142	&	0.02057 \\
55129.71232	& 13.29868	& 0.00148	& 3.62406&	24.69537& 	-12.85689		& 1.27998	&	0.02025 \\
55132.80321	& 13.29327	& 0.00165	& 3.60603&	24.90897& 	-11.99953		& 1.21409	&	0.01984 \\
55135.70052	& 13.28911	& 0.00159	& 3.59773&	24.87190& 	-7.96682		& 1.18975	&	0.02030 \\
55137.70695	& 13.28387	& 0.00136	& 3.60094&	24.84526& 	-13.45768		& 1.18268	&	0.01970 \\
55139.73873	& 13.28154	& 0.00241	& 3.58700&	25.11331& 	-17.36773		& 1.11593	&	0.01968 \\
55141.71517	& 13.28093	& 0.00179	& 3.60334&	24.87551& 	-5.67922		& 1.04440	&	0.01961 \\
55143.68335	& 13.28173	& 0.00193	& 3.60264&	24.91751& 	-7.52072		& 1.32279	&	0.02073 \\
55168.59826	& 13.28915	& 0.00157	& 3.59427&	24.86074& 	-5.09714		& 1.23032	&	0.02106 \\
55169.62393	& 13.29108	& 0.00154	& 3.59996&	25.14133& 	10.91218		& 1.29228	&	0.02133 \\
55230.60800	& 13.29104	& 0.00190	& 3.61222&	25.09914& 	-7.36077		& 1.31543	&	0.01965 \\
55403.92062	& 13.29672	& 0.00278	& 3.58700&	25.11331& 	-17.36773		& 1.59816	&	0.02079 \\
55411.88037	& 13.29203	& 0.00522	& 3.59664&	24.89766& 	-0.57440		& 1.23556	&	0.01964 \\
55428.88535	& 13.30621	& 0.00220	& 3.59963&	24.74944& 	-10.13777		& 1.44940	&	0.02034 \\
55437.86529	& 13.30170	& 0.00148	& 3.60730&	25.03872& 	-8.35145		& 1.28391	&	0.02120 \\
55444.82380	& 13.29678	& 0.00165	& 3.61464&	25.10732& 	-14.16024		& 1.19867	&	0.01971 \\
55450.88896	& 13.29711	& 0.00168	& 3.58700&	25.11331& 	-17.36773		& 1.28313	&	0.02118 \\
55453.89333	& 13.29977	& 0.00172	& 3.59617&	24.94721& 	-1.72937		& 1.14084	&	0.01986 \\
55454.82171	& 13.30149	& 0.00173	& 3.60655&	24.93891& 	-3.71860		& 1.17555	&	0.01868 \\
55455.84630	& 13.30279	& 0.00240	& 3.60733&	24.96234& 	-7.36077		& 1.44517	&	0.02060 \\
55456.83313	& 13.29964	& 0.00158	& 3.60334&	24.87551& 	-5.67922		& 1.16269	&	0.02018 \\
55457.86351	& 13.30165	& 0.00171	& 3.61401&	24.96484& 	-10.77753		& 1.14233	&	0.02044 \\
55493.82107	& 13.30899	& 0.00183	& 3.60161&	24.82933& 	-16.40298		& 1.01267	&	0.01915 \\
55494.76977	& 13.30460	& 0.00201	& 3.60531&	24.80336& 	-19.38256		& 0.90454	&	0.01979 \\
55495.88051	& 13.30289	& 0.00147	& 3.60443&	25.03226& 	-13.75953		& 1.15078	&	0.02096 \\
55497.75076	& 13.29815	& 0.00172	& 3.59634&	24.89247& 	-8.31935		& 1.45566	&	0.01991 \\
55500.80695	& 13.29347	& 0.00177	& 3.59427&	24.86074& 	-5.09714		& 1.08966	&	0.01979 \\
55501.75111	& 13.29021	& 0.00164	& 3.60512&	25.07722& 	-3.29884		& 1.20944	&	0.01980 \\
55505.64461	& 13.28387	& 0.00192	& 3.61087&	24.90341& 	-7.36429		& 1.05933	&	0.01954 \\
55514.68731	& 13.28270	& 0.00160	& 3.62495&	25.00670& 	-5.11962		& 1.29677	&	0.02100 \\
55521.72610	& 13.29585	& 0.00176	& 3.59152&	24.95034& 	-5.09714		& 0.99529	&	0.01948 \\
55523.71392	& 13.29791	& 0.00152	& 3.59773&	24.87190& 	-7.96682		& 1.07141	&	0.01796 \\
55547.63328	& 13.29735	& 0.00161	& 3.60982&	25.12058& 	-14.86438		& 1.25905	&	0.02011 \\
55548.59915	& 13.29992	& 0.00209	& 3.60418&	24.92648& 	-6.98776		& 1.10584	&	0.01939 \\
55549.69948	& 13.30091	& 0.00146	& 3.59773&	24.87190& 	-7.96682		& 1.26415	&	0.01951 \\
55576.59973	& 13.29605	& 0.00200	& 3.60971&	24.95140& 	-7.96682		& 1.36858	&	0.02080 \\
55579.69290	& 13.30053	& 0.00174	& 3.60157&	24.88640& 	-7.42069		& 1.26789	&	0.02069 \\
55586.57901	& 13.31199	& 0.00171	& 3.60418&	24.92648& 	-6.98776		& 1.41550	&	0.02043 \\
55612.54944	& 13.30369	& 0.00163	& 3.60987&	24.95725& 	-2.58211		& 0.95858	&	0.02007 \\
55615.50571	& 13.30480	& 0.00136	& 3.59152&	24.95034& 	-5.09714		& 1.34022	&	0.02348 \\
55621.49969	& 13.29348	& 0.00155	& 3.61108&	25.00144& 	-14.68316		& 1.04788	&	0.02114 \\
55817.89452	& 13.29597	& 0.00200	& 3.60593&	24.70919& 	-13.07587		& 1.31675	&	0.02093 \\
55822.87888	& 13.30537	& 0.00148	& 3.59427&	24.86074& 	-5.09714		& 1.08820	&	0.02093 \\
55825.89910	& 13.30804	& 0.00158	& 3.60626&	24.86616& 	-11.95861		& 1.28318	&	0.02123 \\
55828.89202	& 13.30764	& 0.00148	& 3.61087&	24.90341& 	-7.36429		& 1.21319	&	0.02042 \\
55829.85421	& 13.30775	& 0.00158	& 3.61053&	24.63378& 	-13.75953		& 1.03953	&	0.01991 \\
55830.87234	& 13.30798	& 0.00174	& 3.61108&	25.01686& 	-12.34296		& 1.05729	&	0.02050 \\
55831.83366	& 13.30569	& 0.00145	& 3.59617&	24.94721& 	-1.72937		& 1.16617	&	0.02043 \\
55834.83622	& 13.30246	& 0.00178	& 3.60731&	24.76752& 	-6.97071		& 0.90912	&	0.01919 \\
55835.76068	& 13.30133	& 0.00179	& 3.60731&	24.76752& 	-6.97071		& 1.04509	&	0.01879 \\
55837.83127	& 13.30134	& 0.00134	& 3.61401&	24.96484& 	-10.77753		& 0.99007	&	0.01913 \\
55839.82111	& 13.29755	& 0.00191	& 3.61053&	24.63378& 	-13.75953		& 1.24796	&	0.01921 \\
55840.87205	& 13.29215	& 0.00154	& 3.60593&	24.70919& 	-13.07587		& 1.15323	&	0.01961 \\
55841.75022	& 13.29021	& 0.00209	& 3.60443&	25.03226& 	-13.75953		& 1.12326	&	0.01967 \\
55843.85888	& 13.28824	& 0.00153	& 3.63220&	24.65556& 	5.80814         	& 1.15736	&	0.01989 \\
55870.76889	& 13.28400	& 0.00177	& 3.60512&	25.07722& 	-3.29884		& 1.18192	&	0.02037 \\
55887.68502	& 13.29405	& 0.00264	& 3.60475&	24.90774& 	-14.41329		& 1.31407	&	0.02025 \\
55890.67297	& 13.30346	& 0.00155	& 3.61108&	25.00144& 	-14.68316		& 1.32079	&	0.02020 \\
55893.65958	& 13.30051	& 0.00181	& 3.60987&	24.95725& 	-2.58211		& 1.33008	&	0.02026 \\
55924.64545	& 13.30390	& 0.00173	& 3.59588&	24.67017& 	-11.45474		& 1.23348	&	0.01990 \\
55925.65007	& 13.30571	& 0.00170	& 3.59588&	24.67017& 	-11.45474		& 1.05004	&	0.01901 \\
55926.63066	& 13.30337	& 0.00161	& 3.60593&	24.70919& 	-13.07587		& 1.34434	&	0.02102 \\
55927.63859	& 13.30422	& 0.00199	& 3.59468&	24.91635& 	-6.84679		& 1.13228	&	0.02013 \\
55928.68246	& 13.30283	& 0.00183	& 3.60414&	24.99914& 	-11.90708		& 1.02394	&	0.01955 \\
55929.61470	& 13.30281	& 0.00182	& 3.59996&	25.14133& 	10.91218		& 1.06086	&	0.01966 \\
55930.63306	& 13.30300	& 0.00180	& 3.60094&	24.84526& 	-13.45768		& 1.32293	&	0.02008 \\
55931.61582	& 13.30170	& 0.00194	& 3.60673&	24.81264& 	-13.23527		& 0.95326	&	0.01945 \\
55932.61626	& 13.29942	& 0.00163	& 3.60836&	25.01811& 	-11.99953		& 1.26315	&	0.01968 \\
55933.63090	& 13.30062	& 0.00164	& 3.60987&	24.95725& 	-2.58211		& 1.28021	&	0.01986 \\
55940.55035	& 13.28895	& 0.00246	& 3.60512&	25.07722& 	-3.29884		& 1.54123	&	0.02028 \\
55941.64214	& 13.29572	& 0.00152	& 3.60512&	25.07722& 	-3.29884		& 1.34513	&	0.01968 \\
55942.69832	& 13.29324	& 0.00199	& 3.59387&	24.89050& 	-9.27864		& 0.78451	&	0.01891 \\
55943.63946	& 13.29661	& 0.00209	& 3.59159&	25.00795& 	-14.86627		& 1.39953	&	0.02046 \\
55944.63016	& 13.29985	& 0.00194	& 3.60512&	25.07722& 	-3.29884		& 1.27652	&	0.01975 \\
55945.64138	& 13.29997	& 0.00175	& 3.60836&	25.01811& 	-11.99953		& 1.12134	&	0.02054 \\
55946.63701	& 13.29891	& 0.00172	& 3.61087&	24.90341& 	-7.36429		& 1.19637	&	0.02023 \\
55947.62134	& 13.29747	& 0.00162	& 3.60847&	24.79972& 	-14.86627		& 1.11409	&	0.01987 \\
55949.61627	& 13.30076	& 0.00160	& 3.60593&	24.70919& 	-13.07587		& 1.27358	&	0.01952 \\
55950.62808	& 13.30303	& 0.00170	& 3.59427&	24.86074& 	-5.09714		& 1.40957	&	0.02118 \\
55997.53126	& 13.27822	& 0.00210	& 3.59661&	24.78567& 	-0.57440		& 1.19986	&	0.01991 \\
56001.48895	& 13.27508	& 0.00181	& 3.59152&	24.95034& 	-5.09714		& 1.11461	&	0.01953 \\
56008.50764	& 13.28963	& 0.00166	& 3.60475&	24.90774& 	-14.41329		& 0.91820	&	0.02015 \\
56010.51342	& 13.29753	& 0.00169	& 3.60687&	24.94252& 	-14.61256		& 1.23211	&	0.01990 \\
56022.50593	& 13.29470	& 0.00233	& 3.60414&	24.99914& 	-11.90708		& 1.26817	&	0.02117 \\
56025.48481	& 13.28588	& 0.00262	& 3.61057&	24.94123& 	-15.05656		& 1.71769	&	0.02326 \\
56030.48582	& 13.27992	& 0.00314	& 3.62180&	24.78081& 	-3.29884		& 1.08995	&	0.02079 \\
56032.47450	& 13.28219	& 0.00329	& 3.61087&	24.90341& 	-7.36429		& 0.94920	&	0.02031 \\
56158.86499	& 13.29079	& 0.00207	& 3.61087&	24.90341& 	-7.36429		& 1.01209	&	0.02149 \\
56160.88634	& 13.29507	& 0.00184	& 3.60945&	24.54870& 	-7.36429		& 1.32960	&	0.02247 \\
56171.84600	& 13.30300	& 0.00173	& 3.61245&	24.91672& 	-4.94472		& 1.04735	&	0.01955 \\
56208.85566	& 13.29100	& 0.00315	& 3.60161&	24.82933& 	-16.40298		& 1.55485	&	0.02022 \\
56209.86400	& 13.28648	& 0.00395	& 3.60161&	24.82933& 	-16.40298		& 1.43810	&	0.01896 \\
56210.84941	& 13.29442	& 0.00188	& 3.61586&	24.87951& 	-11.28642		& 1.14180	&	0.01895 \\
56221.86931	& 13.30292	& 0.00222	& 3.60731&	24.76752& 	-6.97071		& 1.17186	&	0.02089 \\
56229.75711	& 13.30168	& 0.00175	& 3.60945&	24.54870& 	-7.36429		& 1.20497	&	0.02203 \\
56230.70914	& 13.30124	& 0.00215	& 3.60945&	24.54870& 	-7.36429		& 1.27728	&	0.02073 \\
56231.78090	& 13.29747	& 0.00171	& 3.60414&	24.99914& 	-11.90708		& 1.22611	&	0.02255 \\
56235.66439	& 13.29494	& 0.00245	& 3.60094&	24.84526& 	-13.45768		& 1.24985	&	0.02070 \\
56236.65314	& 13.29105	& 0.00225	& 3.60161&	24.82933& 	-16.40298		& 1.19125	&	0.02073 \\
56237.64360	& 13.28785	& 0.00194	& 3.60687&	24.94252& 	-14.61256		& 0.99914	&	0.01952 \\
56238.61253	& 13.28867	& 0.00191	& 3.61290&	24.88298& 	-16.40298		& 1.03800	&	0.02031 \\
56239.66651	& 13.28278	& 0.00229	& 3.59152&	24.95034& 	-5.09714		& 1.39054	&	0.02107 \\
56245.61339	& 13.28206	& 0.00229	& 3.60094&	24.84526& 	-13.45768		& 1.08137	&	0.02135 \\
56248.64851	& 13.28634	& 0.00250	& 3.59152&	24.95034& 	-5.09714		& 1.04436	&	0.01965 \\
56249.67253	& 13.28603	& 0.00203	& 3.61977&	24.86813& 	-13.07587		& 1.04205	&	0.01879 \\
56251.69011	& 13.29099	& 0.00164	& 3.61977&	24.86813& 	-13.07587		& 0.97448	&	0.01930 \\
56252.65403	& 13.29243	& 0.00161	& 3.59427&	24.86074& 	-5.09714		& 0.90649	&	0.02016 \\
56253.65586	& 13.29557	& 0.00207	& 3.61057&	24.94123& 	-15.05656		& 1.04962	&	0.01978 \\
56256.68998	& 13.30221	& 0.00199	& 3.61028&	25.10494& 	-15.05656		& 0.91982	&	0.01981 \\
56257.72995	& 13.29963	& 0.00168	& 3.59427&	24.86074& 	-5.09714		& 1.14657	&	0.02025 \\
56259.64344	& 13.30642	& 0.00194	& 3.59427&	24.86074& 	-5.09714		& 1.25647	&	0.01963 \\
56263.63688	& 13.30049	& 0.00252	& 3.61977&	24.86813& 	-13.07587		& 1.26143	&	0.02204 \\
56264.78151	& 13.29736	& 0.00144	& 3.61057&	24.94123& 	-15.05656		& 1.41208	&	0.02171 \\
56283.56533	& 13.29611	& 0.00203	& 3.60761&	24.95206& 	-7.97171		& 1.07310	&	0.02025 \\
56304.65270	& 13.29140	& 0.00186	& 3.61977&	24.86813& 	-13.07587		& 1.22813	&	0.02022 \\
56307.60791	& 13.29589	& 0.00154	& 3.60660&	24.80012& 	-14.61256		& 1.22408	&	0.02136 \\
56312.60633	& 13.30676	& 0.00191	& 3.60660&	24.80012& 	-14.61256		& 1.24584	&	0.02156 \\
56314.60875	& 13.30923	& 0.00197	& 3.60660&	24.80012& 	-14.61256		& 1.19865	&	0.02082 \\
56316.64222	& 13.31724	& 0.00210	& 3.60687&	24.94252& 	-14.61256		& 0.84569	&	0.02116 \\
56318.64452	& 13.31640	& 0.00228	& 3.59152&	24.95034& 	-5.09714		& 1.25696	&	0.02065 \\
56319.65567	& 13.31447	& 0.00249	& 3.59152&	24.95034& 	-5.09714		& 1.13835	&	0.02063 \\
\end{longtable}
}

\longtab{2}{
\begin{longtable}{llllllll}
\caption{GJ 3341 RVs, its uncertainty and activity indicators.}\\
\hline\hline
BJD - 2400000 & RV &  $\sigma_{RV}$ & FWHM  & Contrast & BIS  & S-index & $H\alpha$ \\
 & $[kms^{-1}]$ & $[kms^{-1}]$ & [$kms^{-1}$] & & [$kms^{-1}$] &&  \\
\hline
\endfirsthead
\caption{continued.}\\
\hline\hline
BJD - 2400000 & RV &  $\sigma_{RV}$ & FWHM  & Contrast & BIS  & S-index & $H\alpha$ \\
 & $[kms^{-1}]$ & $[kms^{-1}]$ & [$kms^{-1}$] & & [$kms^{-1}$] &&  \\
\hline
\endhead
\hline
\endfoot
54807.69849	& 47.79778	& 0.00448	& 3.59435	& 25.43663	& -1.56577	& 0.26432	& 0.01753 \\
54825.66746	& 47.80633	& 0.00139	& 3.59007	& 24.79779	& 4.10035	& 0.98919	& 0.01763 \\
54826.64838	& 47.80369	& 0.00166	& 3.59320	& 24.70898	& -14.54631	& 0.94571	& 0.01765 \\
54827.66095	& 47.80152	& 0.00156	& 3.58260	& 24.76051	& -3.80966	& 0.98700	& 0.01791 \\
54828.67307	& 47.80220	& 0.00184	& 3.60239	& 24.53064	& 1.53949	& 0.95732	& 0.01829 \\
54829.78186	& 47.80221	& 0.00182	& 3.59904	& 24.80998	& -10.57797	& 1.06308	& 0.01742 \\
54830.65130	& 47.79394	& 0.00191	& 3.60239	& 24.53064	& 1.53949	& 0.80047	& 0.01796 \\
54831.64673	& 47.79584	& 0.00186	& 3.60781	& 24.79868	& -6.38812	& 0.86513	& 0.01774 \\
54832.65658	& 47.79647	& 0.00288	& 3.61297	& 24.63142	& -15.18704	& 0.95837	& 0.01823 \\
54833.64792	& 47.80005	& 0.00237	& 3.58617	& 24.68657	& -4.75891	& 0.98841	& 0.01869 \\
54834.77493	& 47.79813	& 0.00196	& 3.58447	& 24.62698	& -9.99006	& 0.89096	& 0.01775 \\
54849.56753	& 47.80253	& 0.00133	& 3.60043	& 24.95740	& -7.82953	& 0.76761	& 0.01751 \\
54851.62802	& 47.80455	& 0.00163	& 3.59861	& 24.84315	& -2.13597	& 0.86549	& 0.01709 \\
54881.63742	& 47.80599	& 0.00185	& 3.60168	& 24.77744	& -7.82953	& 0.88508	& 0.01763 \\
54882.59568	& 47.80655	& 0.00152	& 3.57915	& 24.67109	& -15.46011	& 0.87898	& 0.01785 \\
54885.54247	& 47.79999	& 0.00149	& 3.59581	& 24.69552	& -16.07110	& 0.72763	& 0.01762 \\
54932.48536	& 47.80336	& 0.00141	& 3.59952	& 24.69259	& -10.90757	& 0.98036	& 0.01781 \\
54934.48545	& 47.80111	& 0.00183	& 3.57877	& 24.88057	& -9.09266	& 0.93246	& 0.01746 \\
54937.48457	& 47.80246	& 0.00171	& 3.59007	& 24.79779	& 4.10035	& 0.71549	& 0.01790 \\
54940.49248	& 47.80722	& 0.00198	& 3.59435	& 25.43663	& -1.56577	& 0.75937	& 0.01792 \\
54941.48780	& 47.80597	& 0.00146	& 3.59595	& 24.58150	& -7.28340	& 0.78320	& 0.01771 \\
54949.46243	& 47.80247	& 0.00164	& 3.60630	& 24.90630	& -22.68062	& 0.89015	& 0.01784 \\
54950.46139	& 47.80259	& 0.00148	& 3.60283	& 24.58144	& -4.45534	& 0.83795	& 0.01809 \\
54954.46978	& 47.80861	& 0.00174	& 3.58580	& 24.60535	& 2.06906	& 0.52661	& 0.01774 \\
54955.46854	& 47.80541	& 0.00149	& 3.58382	& 24.68647	& -10.98534	& 0.93699	& 0.01811 \\
55220.56224	& 47.80338	& 0.00285	& 3.61297	& 24.63142	& -15.18704	& 0.81790	& 0.01807 \\
55225.54498	& 47.80642	& 0.00189	& 3.57878	& 24.65019	& -12.22348	& 0.78821	& 0.01801 \\
55229.69119	& 47.79952	& 0.00419	& 3.59055	& 24.71588	& -14.52515	& 1.01882	& 0.01853 \\
55423.91008	& 47.81054	& 0.00241	& 3.60986	& 24.65434	& -15.52187	& 0.73691	& 0.01887 \\
55425.90482	& 47.80960	& 0.00193	& 3.59595	& 24.58150	& -7.28340	& 0.97592	& 0.01956 \\
55427.92239	& 47.80106	& 0.00192	& 3.60196	& 24.58954	& -4.45534	& 1.00509	& 0.01805 \\
55428.90390	& 47.80268	& 0.00221	& 3.60507	& 24.73606	& -15.18704	& 0.72720	& 0.01838 \\
55434.88752	& 47.80564	& 0.00188	& 3.60725	& 24.58705	& -12.05471	& 0.89913	& 0.01830 \\
55437.84917	& 47.80258	& 0.00159	& 3.58580	& 24.60535	& 2.06906	&0.79242	& 0.01857 \\
55444.85801	& 47.79774	& 0.00174	& 3.58447	& 24.62698	& -9.99006	& 0.79058	& 0.01831 \\
55480.81601	& 47.80086	& 0.00200	& 3.59904	& 24.80998	& -10.57797	& 0.35148	& 0.01846 \\
55483.80314	& 47.79988	& 0.00203	& 3.59435	& 25.43663	& -1.56577	& 0.86845	& 0.01922 \\
55486.76213	& 47.79658	& 0.00202	& 3.61114	& 24.58479	& -17.42168	& 0.86073	& 0.01834 \\
55492.87155	& 47.80689	& 0.00204	& 3.60507	& 24.73606	& -15.18704	& 0.23161	& 0.01807 \\
55495.89290	& 47.80079	& 0.00141	& 3.57375	& 25.51735	& -18.68856	& 0.45954	& 0.01823 \\
55497.80927	& 47.80042	& 0.00188	& 3.59822	& 24.88792	& -6.20478	& 0.62883	& 0.01827 \\
55499.87347	& 47.80078	& 0.00237	& 3.60630	& 24.90630	& -22.68062	& 0.57420	& 0.01850 \\
55501.81020	& 47.80021	& 0.00196	& 3.58758	& 24.75806	& -12.22348	& 0.39399	& 0.01906 \\
55505.70282	& 47.79806	& 0.00182	& 3.61193	& 24.76277	& -10.33153	& 0.33693	& 0.01929 \\
55506.78000	& 47.79756	& 0.00168	& 3.59007	& 24.79779	& 4.10035	& 0.64716	& 0.01909 \\
55509.76158	& 47.80096	& 0.00158	& 3.58447	& 24.62698	& -9.99006	& 0.77938	& 0.01901 \\
55510.73594	& 47.79845	& 0.00152	& 3.58758	& 24.75806	& -12.22348	& 0.64301	& 0.01899 \\
55512.72825	& 47.79901	& 0.00152	& 3.61448	& 24.53406	& 2.04596	& 0.69542	& 0.01874 \\
55513.74985	& 47.80232	& 0.00216	& 3.60000	& 24.56756	& -13.98973	& 0.71029	& 0.01847 \\
55514.75421	& 47.79813	& 0.00163	& 3.61448	& 24.53406	& 2.04596	& 0.72800	& 0.01848 \\
55518.69233	& 47.80377	& 0.00215	& 3.60202	& 24.58428	& -12.22297	& 0.85978	& 0.01798 \\
55519.63904	& 47.80124	& 0.00220	& 3.58447	& 24.62698	& -9.99006	& 0.77323	& 0.01873 \\
55521.74963	& 47.80237	& 0.00197	& 3.60000	& 24.56756	& -13.98973	& 0.67096	& 0.01865 \\
55522.70507	& 47.80259	& 0.00195	& 3.60158	& 24.22742	& -9.77163	& 0.74732	& 0.01857 \\
55523.75227	& 47.80495	& 0.00197	& 3.58240	& 24.93187	& -12.58244	& 0.80502	& 0.01825 \\
55539.63328	& 47.79842	& 0.00195	& 3.59768	& 24.61102	& -10.57797	& 0.57500	& 0.01836 \\
55542.69632	& 47.79789	& 0.00193	& 3.59055	& 24.71588	& -14.52515	& 0.73343	& 0.01846 \\
55544.69892	& 47.80078	& 0.00146	& 3.61297	& 24.63142	& -15.18704	& 0.64341	& 0.01848 \\
55546.60740	& 47.79905	& 0.00183	& 3.58382	& 24.68647	& -10.98534	& 0.83876	& 0.01826 \\
55547.68313	& 47.80689	& 0.00181	& 3.61073	& 24.59018	& -9.27766	& 0.71074	& 0.01818 \\
55817.91683	& 47.80415	& 0.00163	& 3.58758	& 24.75806	& -12.22348	& 0.69149	& 0.01819 \\
55871.72296	& 47.80264	& 0.00189	& 3.56078	& 24.55298	& -2.50354	& 0.76060	& 0.01884 \\
55875.75986	& 47.80443	& 0.00166	& 3.60406	& 24.82217	& -17.61990	& 0.78623	& 0.01802 \\
55887.80195	& 47.80000	& 0.00212	& 3.59476	& 24.55047	& -16.48387	& 0.83975	& 0.01854 \\
55889.67389	& 47.80156	& 0.00212	& 3.59822	& 24.88792	& -6.20478	& 0.85840	& 0.01841 \\
55891.67563	& 47.80750	& 0.00185	& 3.61297	& 24.63142	& -15.18704	& 0.53778	& 0.01831 \\
55893.67046	& 47.80062	& 0.00194	& 3.58916	& 24.51550	& -18.65345	& 0.63131	& 0.01860 \\
55895.71856	& 47.80340	& 0.00225	& 3.56078	& 24.55298	& -2.50354	& 0.95793	& 0.01835 \\
55924.65743	& 47.80118	& 0.00156	& 3.62850	& 24.25183	& -18.65345	& 0.60524	& 0.01839 \\
55925.60322	& 47.79804	& 0.00182	& 3.59435	& 25.43663	& -1.56577	& 1.38852	& 0.01824 \\
55926.68814	& 47.79881	& 0.00182	& 3.59476	& 24.55047	& -16.48387	& 0.63496	& 0.01821 \\
55927.67658	& 47.79824	& 0.00225	& 3.59061	& 24.89397	& -17.78810	& 0.93117	& 0.01801 \\
55928.78900	& 47.79444	& 0.00275	& 3.62850	& 24.25183	& -18.65345	& 0.38422	& 0.01788 \\
55929.67904	& 47.80136	& 0.00209	& 3.59492	& 24.50066	& -8.11366	& 0.69748	& 0.01819 \\
55930.68755	& 47.80549	& 0.00375	& 3.59861	& 24.84315	& -2.13597	& 0.89780	& 0.01832 \\
55931.63830	& 47.80153	& 0.00183	& 3.59435	& 25.43663	& -1.56577	& 0.73302	& 0.01842 \\
55932.66330	& 47.80040	& 0.00153	& 3.59597	& 24.72643	& -16.07110	& 0.67666	& 0.01894 \\
55933.70230	& 47.80510	& 0.00194	& 3.59435	& 25.43663	& -1.56577	& 0.69418	& 0.01847 \\
55940.66160	& 47.80219	& 0.00245	& 3.60240	& 24.72910	& -16.86121	& 0.65311	& 0.01811 \\
55941.73303	& 47.80174	& 0.00192	& 3.60202	& 24.58428	& -12.22297	& 0.53807	& 0.01808 \\
55942.65135	& 47.80209	& 0.00165	& 3.58580	& 24.60535	& 2.06906	& 0.83814	& 0.01792 \\
55943.74352	& 47.80267	& 0.00187	& 3.60196	& 24.58954	& -4.45534	& 0.89701	& 0.01786 \\
55944.74860	& 47.80799	& 0.00198	& 3.57878	& 24.65019	& -12.22348	& 0.88405	& 0.01830 \\
55945.74749	& 47.80385	& 0.00201	& 3.59861	& 24.84315	& -2.13597	& 1.17721	& 0.01744 \\
55946.74356	& 47.80763	& 0.00196	& 3.61448	& 24.53406	& 2.04596	& 0.48098	& 0.01728 \\
55947.72814	& 47.80986	& 0.00175	& 3.60507	& 24.73606	& -15.18704	& 0.47686	& 0.01787 \\
55948.72736	& 47.80782	& 0.00182	& 3.59435	& 25.43663	& -1.56577	& 0.67393	& 0.01777 \\
55949.72113	& 47.80933	& 0.00203	& 3.60071	& 24.82345	& -11.87951	& 0.70198	& 0.01790 \\
55950.73282	& 47.80362	& 0.00199	& 3.59822	& 24.88792	& -6.20478	& 0.89344	& 0.01829 \\
55997.54782	& 47.80145	& 0.00223	& 3.59899	& 24.73087	& -2.50354	& 0.77080	& 0.01859 \\
55999.50620	& 47.80363	& 0.00170	& 3.60071	& 24.82345	& -11.87951	& 0.99767	& 0.01797 \\
56001.57468	& 47.80602	& 0.00235	& 3.59061	& 24.89397	& -17.78810	& 1.01487	& 0.01874 \\
56004.48819	& 47.80754	& 0.00181	& 3.58120	& 24.91090	& -18.26073	& 0.81673	& 0.01861 \\
56006.56154	& 47.80192	& 0.00206	& 3.60283	& 24.58144	& -4.45534	& 0.85736	& 0.01833 \\
56008.52020	& 47.79842	& 0.00174	& 3.60406	& 24.82217	& -17.61990	& 0.95240	& 0.01943 \\
56013.54468	& 47.79717	& 0.00215	& 3.59818	& 24.55765	& -14.39398	& 0.98921	& 0.01809 \\
56021.50098	& 47.80481	& 0.00233	& 3.60196	& 24.58954	& -4.45534	& 0.78758	& 0.01770 \\
56024.48555	& 47.80081	& 0.00165	& 3.58729	& 24.69291	& -16.86121	& 0.49211	& 0.01779 \\
56025.49723	& 47.80600	& 0.00236	& 3.59492	& 24.50066	& -8.11366	& 1.52699	& 0.01755 \\
56026.50327	& 47.80542	& 0.00258	& 3.57878	& 24.65019	& -12.22348	& 0.65928	& 0.01765 \\
56029.52945	& 47.79952	& 0.00241	& 3.58729	& 24.69291	& -16.86121	& 0.43675	& 0.01832 \\
56030.51172	& 47.80770	& 0.00292	& 3.58758	& 24.75806	& -12.22348	& 0.35110	& 0.01770 \\
56031.51835	& 47.80923	& 0.00450	& 3.59858	& 24.52790	& -15.87937	& -0.37144      & 0.01787 \\ 
56180.85874	& 47.80221	& 0.00287	& 3.60353	& 24.35406	& -12.44804	& 0.91964	& 0.01849 \\
56181.87228	& 47.80618	& 0.00209	& 3.60071	& 24.82345	& -11.87951	& 1.07214	& 0.01850 \\
56182.87143	& 47.79806	& 0.00299	& 3.58335	& 24.87303	& -8.50981	& 0.36638	& 0.01867 \\
56186.84912	& 47.80768	& 0.00171	& 3.59858	& 24.52790	& -15.87937	& 1.00343	& 0.01910 \\
56187.86191	& 47.80731	& 0.00534	& 3.59861	& 24.84315	& -2.13597	& 1.51751	& 0.01896 \\
56190.88488	& 47.80761	& 0.00226	& 3.59858	& 24.52790	& -15.87937	& 0.85908	& 0.01906 \\
56192.86732	& 47.80342	& 0.00272	& 3.59595	& 24.58150	& -7.28340	& 0.88605	& 0.01926 \\
56193.85343	& 47.80615	& 0.00204	& 3.59822	& 24.88792	& -6.20478	& 1.09271	& 0.01896 \\
56194.87392	& 47.80390	& 0.00214	& 3.59061	& 24.89397	& -17.78810	& 0.69008	& 0.01876 \\
56195.83731	& 47.80180	& 0.00211	& 3.58335	& 24.87303	& -8.50981	& 0.79571	& 0.01870 \\
56196.84078	& 47.80332	& 0.00182	& 3.60071	& 24.82345	& -11.87951	& 0.93673	& 0.01915 \\
56198.85435	& 47.80369	& 0.00359	& 3.61448	& 24.53406	& 2.04596	& 0.03357	& 0.01881 \\
56199.85744	& 47.80601	& 0.00299	& 3.60353	& 24.35406	& -12.44804	& 0.48219	& 0.01979 \\
56200.82093	& 47.80271	& 0.00239	& 3.61448	& 24.53406	& 2.04596	& 0.95819	& 0.02087 \\
56201.82289	& 47.80065	& 0.00227	& 3.57375	& 25.51735	& -18.68856	& 1.16869	& 0.01962 \\
56202.87283	& 47.80820	& 0.00231	& 3.59299	& 24.63871	& -9.11403	& 0.92122	& 0.01907 \\
56235.67528	& 47.80431	& 0.00229	& 3.61565	& 24.52854	& -11.18810	& 0.67675	& 0.01919 \\
56236.68649	& 47.80250	& 0.00212	& 3.60071	& 24.82345	& -11.87951	& 1.06575	& 0.01932 \\
56237.65447	& 47.79827	& 0.00211	& 3.61565	& 24.52854	& -11.18810	& 0.80598	& 0.01872 \\
56238.71167	& 47.80157	& 0.00205	& 3.59299	& 24.63871	& -9.11403	& 0.84328	& 0.01867 \\
56245.62430	& 47.80771	& 0.00231	& 3.59858	& 24.52790	& -15.87937	& 1.22042	& 0.01838 \\
56247.61995	& 47.80508	& 0.00195	& 3.59858	& 24.52790	& -15.87937	& 0.41762	& 0.01926 \\
56248.70192	& 47.80076	& 0.00303	& 3.61222	& 24.92663	& -2.13597	& 0.72502	& 0.01898 \\
56250.65160	& 47.79364	& 0.00302	& 3.58479	& 24.64203	& -11.87951	& 0.69128	& 0.01851 \\
56251.70851	& 47.79628	& 0.00168	& 3.60071	& 24.82345	& -11.87951	& 0.98762	& 0.01985 \\
56252.66497	& 47.79695	& 0.00169	& 3.59861	& 24.84315	& -2.13597	& 0.91803	& 0.01865 \\
56254.67894	& 47.79673	& 0.00170	& 3.60071	& 24.82345	& -11.87951	& 0.90922	& 0.01847 \\
56255.61817	& 47.80003	& 0.00196	& 3.59861	& 24.84315	& -2.13597	& 0.85057	& 0.01912 \\
56258.65243	& 47.80533	& 0.00196	& 3.61222	& 24.92663	& -2.13597	& 0.99596	& 0.01860 \\
56259.65469	& 47.80498	& 0.00206	& 3.59858	& 24.52790	& -15.87937	& 0.71719	& 0.01842 \\
56262.81151	& 47.80699	& 0.00187	& 3.61222	& 24.92663	& -2.13597	& 0.97329	& 0.01809 \\
56263.64804	& 47.80253	& 0.00287	& 3.59861	& 24.84315	& -2.13597	& 0.60908	& 0.01759 \\
\end{longtable}
}

\longtab{3}{
\begin{longtable}{llllllll}
\caption{GJ 3543 RVs, its uncertainty and activity indicators.}\\
\hline\hline
BJD - 2400000 & RV &  $\sigma_{RV}$ & FWHM  & Contrast & BIS  & S-index & $H\alpha$ \\
 & $[kms^{-1}]$ & $[kms^{-1}]$ & [$kms^{-1}$] & & [$kms^{-1}$] &&  \\
\hline
\endfirsthead
\caption{continued.}\\
\hline\hline
BJD - 2400000 & RV &  $\sigma_{RV}$ & FWHM  & Contrast & BIS  & S-index & $H\alpha$ \\
 & $[kms^{-1}]$ & $[kms^{-1}]$ & [$kms^{-1}$] & & [$kms^{-1}$] &&  \\
\hline
\endhead
\hline
\endfoot
54455.81748	&15.09084	&0.00134	&3.72672	&20.74941	&-7.25495	&2.42133	&0.02606 \\
54825.82158	&15.09106	&0.00106	&3.72954	&20.77381	&-10.18650	&2.56120	&0.02738 \\
54826.80721	&15.09176	&0.00099	&3.72600	&20.51434	&-14.55733	&2.81219	&0.02980 \\
54827.79556	&15.09645	&0.00113	&3.72417	&20.60870	&-8.41420	&2.50082	&0.02735 \\
54828.84278	&15.09488	&0.00102	&3.72244	&20.63083	&-9.31131	&2.44846	&0.02575 \\
54829.86968	&15.09389	&0.00116	&3.72703	&20.57069	&-11.47510	&3.22177	&0.03741 \\
54830.84558	&15.09355	&0.00128	&3.72672	&20.74941	&-7.25495	&2.52963	&0.02766 \\
54831.83059	&15.08899	&0.00134	&3.74139	&20.73712	&-18.36757	&2.36949	&0.02655 \\
54832.82130	&15.09527	&0.00202	&3.72834	&20.68230	&-10.32177	&2.49733	&0.02709 \\
54833.84856	&15.08998	&0.00102	&3.72406	&20.60702	&-5.26544	&2.34166	&0.02562 \\
54834.85241	&15.08970	&0.00117	&3.72600	&20.51434	&-14.55733	&2.34512	&0.02590 \\
55585.79129	&15.09399	&0.00112	&3.73041	&20.77120	&-10.91675	&2.31790	&0.02572 \\
55589.80228	&15.09859	&0.00128	&3.72267	&20.56052	&-10.83070	&3.38598	&0.03712 \\
55593.78829	&15.09057	&0.00114	&3.72497	&20.54240	&-15.83386	&2.32222	&0.02496 \\
55622.60745	&15.09096	&0.00108	&3.71482	&20.57990	&-15.38253	&2.38990	&0.02459 \\
55628.66836	&15.09394	&0.00128	&3.72876	&20.54426	&-10.45739	&2.43931	&0.02579 \\
55629.67559	&15.08887	&0.00113	&3.72244	&20.63083	&-9.31131	&2.38816	&0.02512 \\
55630.71219	&15.09126	&0.00113	&3.72556	&20.54603	&-13.22745	&2.26502	&0.02530 \\
55632.67210	&15.09896	&0.00120	&3.73159	&20.72365	&-10.23569	&2.48187	&0.02686 \\
55633.64839	&15.09842	&0.00106	&3.72956	&20.71123	&-9.74785	&2.22836	&0.02441 \\
55634.65201	&15.09676	&0.00111	&3.73331	&20.43016	&-11.08622	&2.21318	&0.02476 \\
55638.63504	&15.08924	&0.00148	&3.72343	&20.56211	&-8.82476	&2.33684	&0.02473 \\
55639.61274	&15.09100	&0.00108	&3.72834	&20.68230	&-10.32177	&2.07769	&0.02330 \\
55642.61391	&15.09322	&0.00114	&3.72020	&20.23990	&-11.97169	&2.28483	&0.02457 \\
55643.65849	&15.09596	&0.00117	&3.72889	&20.73374	&-13.64324	&2.40455	&0.02488 \\
55644.61310	&15.09842	&0.00116	&3.72672	&20.74941	&-7.25495	&2.29391	&0.02432 \\
55648.62383	&15.08819	&0.00105	&3.73091	&20.59071	&-6.85842	&2.31546	&0.02512 \\
55652.47921	&15.09772	&0.00122	&3.73159	&20.72365	&-10.23569	&2.10686	&0.02395 \\
55654.47899	&15.09558	&0.00112	&3.73555	&20.57637	&-11.12994	&2.20247	&0.02458 \\
55657.47653	&15.09423	&0.00106	&3.73033	&20.64375	&-5.23050	&2.19046	&0.02498 \\
55659.47423	&15.09254	&0.00111	&3.72149	&20.50567	&-12.44366	&2.39384	&0.02658 \\
55663.47053	&15.09659	&0.00159	&3.73030	&20.62977	&-15.41546	&2.30264	&0.02629 \\
55672.58915	&15.09790	&0.00132	&3.72621	&20.59418	&-9.43624	&2.40820	&0.02514 \\
55676.56936	&15.08937	&0.00128	&3.72149	&20.50567	&-12.44366	&2.46696	&0.02610 \\
55677.53431	&15.09222	&0.00116	&3.72684	&20.52647	&-11.47510	&2.42020	&0.02593 \\
55711.48912	&15.09185	&0.00135	&3.72591	&20.59257	&-5.23050	&2.20280	&0.02424 \\
55871.80485	&15.09344	&0.00138	&3.72668	&20.74823	&-12.84963	&2.40549	&0.02600 \\
55874.81344	&15.09026	&0.00213	&3.72822	&20.66578	&-11.31153	&1.82523	&0.02621 \\
55875.83369	&15.08913	&0.00117	&3.72700	&20.55149	&-14.82741	&2.49935	&0.02917 \\
55878.83418	&15.08896	&0.00157	&3.73700	&20.74596	&-13.22745	&1.98883	&0.02502 \\
55880.84096	&15.09154	&0.00225	&3.73700	&20.74596	&-13.22745	&1.89586	&0.02629 \\
55887.84060	&15.09473	&0.00145	&3.72962	&20.59683	&-15.87957	&2.11055	&0.02645 \\
55888.81874	&15.09496	&0.00139	&3.72538	&20.63755	&-11.31153	&2.28485	&0.02968 \\
55889.83853	&15.09467	&0.00128	&3.72729	&20.55800	&-14.51422	&2.24282	&0.02658 \\
55890.85136	&15.09197	&0.00122	&3.72556	&20.54603	&-13.22745	&2.23126	&0.02634 \\
55891.82222	&15.09423	&0.00110	&3.72834	&20.68230	&-10.32177	&2.22775	&0.02657 \\
55892.81128	&15.08969	&0.00123	&3.73033	&20.64375	&-5.23050	&2.19169	&0.02685 \\
55893.82921	&15.08623	&0.00126	&3.72497	&20.54240	&-15.83386	&2.63181	&0.02958 \\
55894.80379	&15.08860	&0.00099	&3.71482	&20.57990	&-15.38253	&2.44420	&0.02793 \\
55996.71895	&15.08933	&0.00129	&3.72834	&20.68230	&-10.32177	&2.20589	&0.02427 \\
55997.66483	&15.09181	&0.00141	&3.72429	&20.76135	&-11.42857	&2.00610	&0.02505 \\
56002.71704	&15.09518	&0.00142	&3.72385	&20.47525	&-13.35688	&2.51458	&0.02621 \\
56003.66997	&15.09791	&0.00154	&3.72703	&20.57069	&-11.47510	&2.11422	&0.02514 \\
56005.69427	&15.09413	&0.00111	&3.72385	&20.47525	&-13.35688	&2.22501	&0.02601 \\
56007.61106	&15.09227	&0.00114	&3.72834	&20.68230	&-10.32177	&2.16732	&0.02622 \\
56008.65231	&15.09501	&0.00126	&3.72406	&20.60702	&-5.26544	&2.43808	&0.02755 \\
56010.58917	&15.09231	&0.00124	&3.71482	&20.57990	&-15.38253	&2.09934	&0.02373 \\
56011.63581	&15.09124	&0.00105	&3.72834	&20.68230	&-10.32177	&2.21262	&0.02530 \\
56012.58228	&15.08975	&0.00108	&3.72668	&20.74823	&-12.84963	&2.13713	&0.02570 \\
56020.62473	&15.09582	&0.00120	&3.72889	&20.73374	&-13.64324	&2.35845	&0.02669 \\
56021.55890	&15.09221	&0.00121	&3.72874	&20.80548	&-12.44366	&2.20731	&0.02567 \\
56022.61059	&15.09188	&0.00136	&3.72429	&20.76135	&-11.42857	&2.29623	&0.02833 \\
56024.59278	&15.09162	&0.00110	&3.72729	&20.55800	&-14.51422	&2.19298	&0.02568 \\
56025.61982	&15.09072	&0.00181	&3.72729	&20.55800	&-14.51422	&2.05005	&0.02601 \\
56026.59209	&15.09309	&0.00124	&3.72874	&20.55294	&-11.95272	&2.54414	&0.02647 \\
56031.65055	&15.09088	&0.00400	&3.71807	&20.53797	&-15.38253	&0.88335	&0.02646 \\
56354.75925	&15.08873	&0.00114	&3.72874	&20.80548	&-12.44366	&2.28217	&0.02744 \\
56355.73982	&15.09588	&0.00320	&3.72516	&20.45809	&-18.36757	&0.79015	&0.02421 \\
56360.76207	&15.09451	&0.00124	&3.73555	&20.57637	&-11.12994	&2.18953	&0.02353 \\
56361.72566	&15.09299	&0.00140	&3.71807	&20.53797	&-15.38253	&1.86769	&0.02306 \\
56363.71853	&15.09338	&0.00120	&3.72889	&20.73374	&-13.64324	&2.04838	&0.02452 \\
56365.66954	&15.09406	&0.00114	&3.72889	&20.73374	&-13.64324	&2.21021	&0.02570 \\
56366.62280	&15.09451	&0.00221	&3.72149	&20.50567	&-12.44366	&2.23726	&0.02479 \\
56367.59345	&15.09417	&0.00126	&3.71482	&20.57990	&-15.38253	&3.05196	&0.03071 \\
56369.64591	&15.09181	&0.00116	&3.72874	&20.55294	&-11.95272	&2.62690	&0.03010 \\
56370.68844	&15.09180	&0.00116	&3.72149	&20.50567	&-12.44366	&3.07099	&0.03141 \\
56371.68821	&15.08913	&0.00121	&3.72149	&20.50567	&-12.44366	&2.26730	&0.02501 \\
56372.63459	&15.08860	&0.00124	&3.75693	&20.44840	&-20.34396	&2.76210	&0.03103 \\
56373.65240	&15.08967	&0.00135	&3.72874	&20.80548	&-12.44366	&2.24918	&0.02510 \\
56374.66466	&15.08518	&0.00113	&3.72149	&20.50567	&-12.44366	&2.12089	&0.02441 \\
\end{longtable}
}

\end{appendix}
\end{document}